\pdfoutput=1

\documentclass[iop]{emulateapj}
\usepackage{color}
\usepackage{hyperref}
\usepackage{ulem}

\usepackage[]{natbib}
\usepackage{amssymb}
\usepackage{bbm}
\usepackage{txfonts}
\usepackage{psfrag}
\usepackage{graphicx}
\usepackage{natbib}
\usepackage{if then}
\usepackage{longtable}
\usepackage{color}
\usepackage{tikz}
\usetikzlibrary{shapes,arrows}
\usepackage{verbatim}

\def\del#1{{}}

\newcommand{\mat}[1]{{#1}}
\newcommand{\newtext}[1]{{{#1}}}

\shorttitle{Methods for Bayesian power spectrum inference with galaxy surveys}
\shortauthors{Jasche et al.}

\begin{document}

\title{Methods for Bayesian power spectrum inference with galaxy surveys}


\author{Jens Jasche\altaffilmark{1}, Benjamin D. Wandelt\altaffilmark{1,2,3}}
\affil{\altaffilmark{1} CNRS, UMR 7095, Institut d'Astrophysique de Paris, 98 bis, boulevard Arago, 75014 Paris, France \\
\altaffilmark{2} UPMC Univ Paris 06, UMR 7095, Institut d'Astrophysique de Paris, 98 bis, boulevard Arago, 75014 Paris, France\\
\altaffilmark{3} Departments of Physics and Astronomy, 1110 W Green Street, University of Illinois at Urbana-Champaign, Urbana, IL 61801, USA\\
}








\begin{abstract}
We derive and implement a full Bayesian large scale structure inference method aiming at precision
recovery of the cosmological power spectrum from galaxy redshift surveys. Our  approach improves over previous Bayesian methods by performing
a joint inference of the three dimensional density field, the cosmological power spectrum, luminosity dependent galaxy biases and corresponding normalizations. We account for all joint and correlated uncertainties between all inferred  quantities. Classes of galaxies with different biases are treated as separate sub samples. The method therefore also allows the combined analysis of more than one galaxy survey.
 In particular, it solves the problem of inferring the power spectrum from galaxy surveys with non-trivial survey geometries  by exploring the joint posterior distribution
with efficient implementations of multiple block Markov chain and Hybrid Monte Carlo methods. Our Markov sampler achieves high statistical efficiency  in low signal to noise regimes by
using a deterministic reversible jump algorithm. This approach reduces the correlation length of the sampler by several orders of magnitude turning the otherwise numerical unfeasible problem of joint parameter exploration into a numerically manageable task.
We test our method on an artificial mock galaxy survey, emulating characteristic features of the Sloan Digital Sky Survey data release 7, such as its survey geometry and luminosity
dependent biases. These tests demonstrate the numerical feasibility of our large scale Bayesian inference frame work when the parameter space has  millions of dimensions.
 The method reveals and correctly treats the anti-correlation between bias amplitudes and power spectrum, which are not taken into account in current approaches to power spectrum estimation, a  20 percent effect across large ranges in k-space.   In addition, the method results in constrained realizations of density fields obtained without assuming the power spectrum or bias parameters in advance. 
\end{abstract}

\keywords{cosmology: observations: methods: numerical}

\section{Introduction}
With the advent of large three dimensional galaxy surveys the three dimensional
large scale structure has become a major source of knowledge to understand the homogeneous and inhomogeneous evolution of our Universe.
In particular, two-point statistics of the three dimensional matter distribution, are important tools to test our current
picture of inflation and evaluate cosmological models describing the origin and evolution of the Universe. For example, detailed knowledge on the overall shape of the matter
power spectrum can provide constraints on neutrino masses or the primordial power spectrum and break degeneracies in cosmological parameter estimation from Cosmic Microwave Background (CMB) data when measuring the parameter combination \(\Omega_m/h\) \citep[e.g.][]{hu-98,wmap-spergel,HANNESTAD2003,Efstathiou_2002,PERCIVAL2002,wmap-spergel,VERDE2003}.
Since the physics governing the baryon acoustic oscillations (BAO) in the power spectrum is well understood \citep[e.g.][]{SILK1968,PEEBLES1970,SUNYAEV1970}, precise measurements of the BAO will allow us to establish this length scale as a new standard ruler to measure the geometry of our Universe through the redshift distance relation \citep[][]{BLAKE2003,SEO2003}. Inferring the BAOs from observation therefore constitutes an important task for determining the detailed nature of a possible dynamic dark energy component, which may explain the
currently observed accelerated expansion of the Universe.

\begin{figure*}
	\centering
	{
	\tikzstyle{blank} = [rectangle,  fill=white!20,text width=5em, text centered, rounded corners, minimum height=4em]
	\tikzstyle{decision} = [diamond, draw, fill=blue!40,text width=4.5em, text badly centered, node distance=3cm, inner sep=0pt]
	\tikzstyle{block} = [rectangle, draw, fill=blue!60,text width=7em, text centered, rounded corners, minimum height=4em]
	\tikzstyle{line} = [draw, -latex']
	\tikzstyle{cloud} = [draw, ellipse,fill=red!40, node distance=3cm, minimum height=2em]
	\tikzstyle{clouda} = [draw, ellipse,fill=green!60, text centered,text width=3em, node distance=3cm, minimum height=2em]
	\tikzstyle{cloudblank} = [rectangle,  fill=white!20,text width=5em, text centered, rounded corners, minimum height=4em]
\begin{tikzpicture}[node distance = 1.5cm, auto]
    \node [clouda] (d1) {data \\ \(d_1\), \(d_2\) };
    \node [block,below of=d1] (DENSSAMPLING) {density sampling \\ \(\delta\)};
    \node [blank,below of=DENSSAMPLING] (blank1) {};
    \node [blank,below of=blank1] (blank2) {};
    \node [blank,below of=blank2] (blank3) {};
    \node [blank,left of=blank2] (blank2a) {};
    \node [blank,right of=blank2] (blank2b) {};
    \node [block,right of=blank2b] (NEMANSAMPLING) { \(\bar{N}\) sampling \\ \(\bar{N}_1\),\(\bar{N}_2\)};
    \node [block,left  of=blank2a] (BIASSAMPLING) {bias sampling\\ \(b1\), \(b2\)};
    \node [block,below of=blank3] (SPECSAMPLING) {power spectrum sampling\\ \(P(k)\)};	
    \node [cloud,left of=blank1] (denssample) {new \(\delta\) sample};
    \node [cloud,right of=blank1] (nmeansample) {new \(\bar{N}\) samples};
    \node [cloud,left of=blank3] (biassample) {new \(b\) samples};
    \node [cloud,right of=blank3] (pspecsample) {new \(P(k)\) sample};
    \node [clouda,right of=NEMANSAMPLING] (d2) {data \\ \(d_1\), \(d_2\) };
    \node [clouda,left of=BIASSAMPLING] (d3) {data \\ \(d_1\), \(d_2\) };
    \path [line] (d2) -- (NEMANSAMPLING);
    \path [line] (d3) -- (BIASSAMPLING);
    \path [line] (d1) -- (DENSSAMPLING);
    \path [line] (DENSSAMPLING) -| (denssample);
    \path [line] (denssample) -- (BIASSAMPLING);
    \path [line] (BIASSAMPLING) -- (biassample);
    \path [line] (biassample) |- (SPECSAMPLING);
    \path [line] (SPECSAMPLING) -| (pspecsample)  ;
    \path [line] (pspecsample) -- (NEMANSAMPLING)  ;
    \path [line] (NEMANSAMPLING) -- (nmeansample)  ;
    \path [line] (nmeansample) |- (DENSSAMPLING);
\end{tikzpicture}
	}
\caption{Flow chart depicting the multi-step iterative block sampling procedure exemplified for two data sets. In the first step a three dimensional density field will be realized conditional on the galaxy observations.
In subsequent steps, the bias parameters, the power spectrum and the normalization parameters for the galaxy distribution will be sampled conditional on the respective previous samples. Iteration
of this process yields samples from the full joint posterior distribution.}
	\label{fig:flowchart}
\end{figure*}
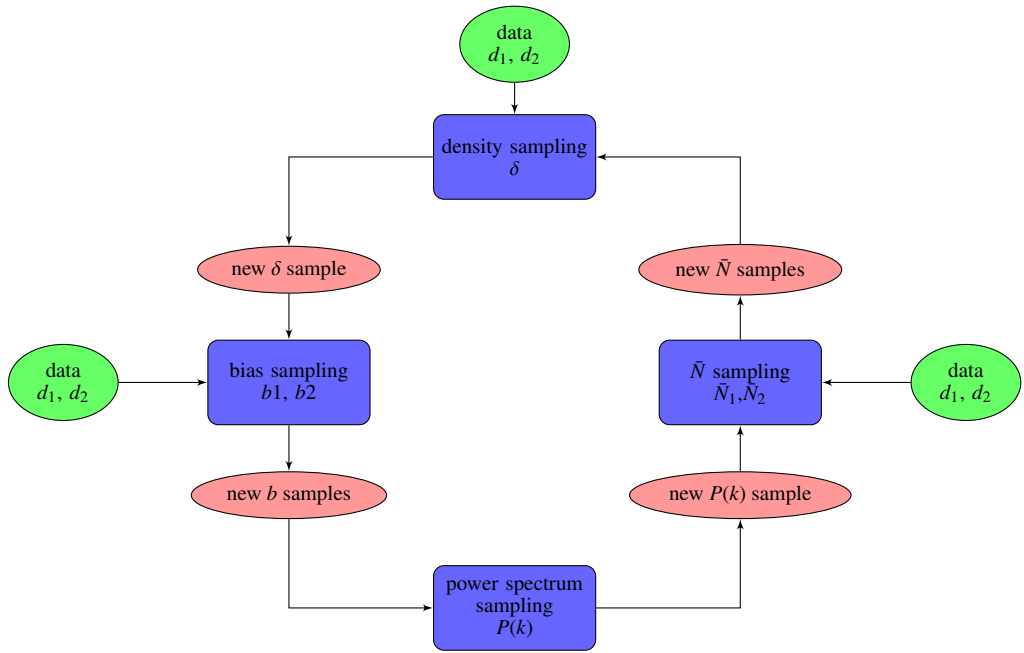
Precision cosmology of this kind requires not only the large and informative data sets which are becoming available, such as the Sloan Digital Sky Survey (SDSS), Dark Energy Survey (DES), Canada-France-Hawaii Telsecope Legacy Survey (CFHTLS) and EUCLID, but also  fair and accurate data analysis methods \citep[][]{YORK2000,COUPON2009,REFREGIER2010,FRIEMAN2011}. Establishing  contact between observations and theoretical predictions in large scale structure is non-trivial. 
This is due to the fact that generally galaxy redshift surveys are subject to a variety of systematic and statistical uncertainties that arise in the observational process such as, noise, survey geometry,  selection effects, and close pair incompleteness due to fiber collisions.

Furthermore, the unknown clustering bias of any one considered galaxy sample compared to the overall mass distribution adds an intrinsic level of  uncertainty \citep[][]{SANCHEZ2008}. Redshift space distortions can be considered a systematic for some analyses but also contain cosmological information.

A careful  treatment of the survey geometry is essential for  the analysis of LSS surveys \citep[][]{TEGMARK1995,BALLINGER1995}.
Generally, the raw power spectrum, estimated from a galaxy redshift survey, yields an expectation value that is the true power spectrum convolved with the survey mask \citep[][]{COLE2005}.
This convolution distorts the shape of the power spectrum and dramatically reduces the detectability of subtle features such as the BAOs.
Also the details of galaxy clustering can have a major impact on the shape of the power spectrum \citep[][]{TEGMARK2004}.
The details of how galaxies cluster and trace the gravitational potentials of dark matter are complicated and not conclusively understood at present.
Most likely the relation between galaxies and matter is essentially non-linear and possibly stochastic in nature, 
such that the inferred galaxy power spectrum is expected to deviate from that of mass \citep[][]{DEKEL1999}.
The issue of the galaxy bias remains complicated even in the limit of a linear bias, since different types of galaxies exhibit different clustering behavior \citep[see e.g.][]{COLE2005}.
As an example, luminosity dependent clustering of galaxies introduces scale dependent biases, when inferring power-spectra from flux limited galaxy surveys \citep[][]{TEGMARK2004}.

\subsection{The state of the art}
To address these problems, a large variety of different power spectrum inference methods have been proposed in the literature.
Substantial work has been done in respect to Fourier transform based methods, such as the optimal weighting scheme.
In this approach galaxies are assigned a weight, in order to reduce the error in the estimated power \citep[see e.g.][]{FELDMAN1994,TEGMARK1995,HAMILTON1997A,YAMAMOTO2003,PERCIVAL2004}.
As alternatives to Fourier space methods, there exist methods relying on Karhunen-Lo\`{e}ve or spherical harmonics decompositions \citep[see e.g.][]{TEGMARK1997,TEGMARK2004,POPE2004,FISHER1994,HEAVENS1995,TADROS1999,PERCIVAL2004,PERCIVAL2005}.
Furthermore, there exists a rich body of research based on a variety of likelihood methods aiming at inferring the real space power spectrum \citep[][]{BALLINGER1995,HAMILTON1997A,HAMILTON1997B,TADROS1999,PERCIVAL2005}.
To not only provide maximum likelihood estimates but also corresponding conditional errors, \cite{PERCIVAL2005} proposed a Markov Chain Monte Carlo approach.

\subsection{Bayesian analysis}
Inference of the three dimensional density field and the corresponding cosmological power spectrum from galaxy redshift surveys with highly structured survey geometries and statistical uncertainties
is an ill-posed problem. This means, there generally exists a large range of feasible solutions consistent with observations. The Bayesian approach solves this problem by providing numerical representations of the joint three dimensional density and power spectrum posterior distribution \citep[][]{JASCHE2010PSPEC}.
Similar approaches have also been applied to Bayesian inference with CMB data \citep[see e.g.][]{WANDELT2004,ODWYER2004,2004ApJS..155..227E,JEWELL2004,LARSON2007,ERIKSEN2007,JEWELL2009}.

In the age of precision cosmology the goal is to assess joint and correlated uncertainties between quantities inferred from a given galaxy redshift survey. A particularly important example are galaxy biases that, if measured incorrectly,  have 
the potential to distort the shape of the inferred power spectrum over a large range of modes \citep{TEGMARK2004}. Traditionally, biases are accounted for in the following  
sequential pipeline process: first, measure the bias amplitudes from power-spectra of different galaxy populations fixing to a fiducial power spectrum; then infer the power spectrum from the entire catalog \citep[see e.g.][]{TEGMARK2004}. This approach leaves open the charge of  overusing the information content of the data because biases and the power spectrum inferences are treated as independent when they are not. 
Alternative approaches like the optimal weighting scheme or the  Bayesian power spectrum inference method
assume fixed bias models \citep[see e.g.][]{PERCIVAL2004,JASCHE2010PSPEC}. 

In all these methods  uncertainties in the inferred bias models do not feed back  to  the power spectrum. It is therefore important to quantify these joint uncertainties and to devise methods to account for these effects.
In this paper we report on the inclusion of these effects in the inference framework described in \citet{JASCHE2010PSPEC}. 

Our approach  accounts for joint uncertainties by exploring the joint posterior distribution of the three dimensional density field, the power spectrum, galaxy biases
and corresponding normalizations conditioned to observations. Numerical and statistical efficient exploration of the corresponding high dimensional parameter space
is achieved via new implementations of multiple block Markov chain and Hybrid Monte Carlo methods. The schematic outline of the iterative block sampling scheme is given in figure \ref{fig:flowchart}. Running through the analysis cycle generates a sampled representation of the full joint posterior distribution, which naturally accounts for joint and correlated uncertainties involved in the inference process.

This paper also describes radical methodological changes in the implementation of the sampling scheme underlying the exploration of this joint posterior density. Using a combination of an efficient Hybrid Monte Carlo sampling method for all the parameters, together with the reversible jump algorithm  described by \citet{JEWELL2009} we obtained an algorithm that proves to sample all the parameters  efficiently both in the high and in the low signal-to-noise regimes.
 
The paper is structured as follows.
In section \ref{Notation}, we summarize the notation used in this work. Section \ref{LSS_POSTERIOR} describes the large scale structure posterior distribution, employed
to infer the large scale three dimensional density field, power spectrum, galaxy biases and normalizations from galaxy observations. Here we also describe how the complex task of
sampling the full joint posterior distribution can be dissected into a sequence of smaller sub problems within the framework of multiple block Metropolis-Hastings sampling.
In Section \ref{numericalimplementation}, we discuss the numerical implementation of our method. The following section \ref{mock_observations} describes the generation of an artificial
galaxy mock observation which emulates characteristic features of the Sloan Digital Sky Survey Data Release 7 \citep[][]{SDSS7}.
In section \ref{TESTING}, we perform a test of our method by applying it to artificial galaxy data. These tests particularly aim at estimating the efficiency of the algorithm in a realistic scenario. 
Some results of the performed test are discussed in section \ref{results}. In particular, here we show, that the posterior correlations between
inferred power-spectra and galaxy biases are strongly anti-correlated over large ranges in Fourier-space.
The paper is concluded in section \ref{Conclusion} by a discussion of the results obtained in this work and some concluding remarks.

\section{Notation}
\label{Notation}
In this section, we describe the basic notation used throughout this work. Let the quantity \(\rho_i=\rho(\vec{x}_i)\) be the field amplitude of the three dimensional field \(\rho(\vec{x})\) at position \(\vec{x}_i\). Then the index \(i\) has to be understood as a multi index, which labels the three components of the position vector:
\begin{equation}
\label{eq:multi_index}
\vec{x}_i =[x^1_i,x^2_i,x^3_i] \, ,
\end{equation}
where \(x^j_i\) is the \(j\)th component of the \(i\)th position vector. Alternatively one can understand the index \(i\) as a set of three indices \(\{r,s,t\}\) so that for an equidistant grid along the three axes the position vector can be expressed as:
\begin{equation}
\label{eq:multi_index_a}
\vec{x}_i =\vec{x}_{r,s,t} = [\Delta x\, r,\Delta y\, s,\Delta z\, t] \, ,
\end{equation}
with \(\Delta x\), \(\Delta y\) and \(\Delta z\) being the grid spacing along the three axes.
With this definition we yield:
\begin{equation}
\label{eq:multi_index_a}
\rho_i \equiv \rho_{r,s,t} \, .
\end{equation}
Also note that any summation running over the multi index \(i\) is defined as the three sums over the three indices \(r\), \(s\) and \(t\):
\begin{equation}
\label{eq:multi_index_c}
\sum_i \equiv \sum_r \sum_s \sum_t \, .
\end{equation}
Further, we will frequently use the notation \(\{\rho_i\}\), which denotes the set of field amplitudes at different positions \(\vec{x}_i\).
In particular:
\begin{equation}
\label{eq:multi_index_set}
\{\rho_i\} \equiv \{\rho_0, \rho_1, \rho_2, ... ,\rho_{N_{vox}-1}\} \, ,
\end{equation}
\newtext{where \(N_{vox}\) is the total number of position vectors or voxels in the three dimensional volume.}

\section{The Large scale structure Posterior}
\label{LSS_POSTERIOR}
In this section we will describe the Large scale structure posterior distribution employed in this work. As will be demonstrated, the multiple block sampling approach
permits to dissect the complex problem of exploring a joint posterior distribution into a set of smaller subtasks.

\subsection{The joint posterior distribution}
The aim of this work is to update and complement the previously presented power spectrum sampling framework described in \citet{JASCHE2010PSPEC} to be numerically more
efficient and also to account for additional systematics and statistical uncertainties of the galaxy distribution from which the density field and the power spectrum
shall be inferred. In particular, the previously developed algorithm accounted for systematics such as arising from survey geometry and selection effects as well
as statistical uncertainties due to noise in the galaxy distribution and cosmic variance \citep[see ][ for details]{JASCHE2010PSPEC}.
Additional uncertainties arise from the galaxy bias and the correct normalization for the galaxy density contrast. Both of these effects can greatly effect the recovery of
the large scale power spectrum and may yield erroneous results when not accounted for accurately. In particular, a galaxy sample consisting of many galaxy types, all tracing
the underlying density field differently, does not provide a homogeneous sample of the underlying density field. This is because luminous galaxies will dominate
the galaxy sample at larger distances while fainter galaxies are found in regions closer to the observer. Due to the different clustering behavior of  
luminous and faint galaxies, this observational effect introduces a spatially varying bias, which translates to a scale depended bias in Fourier-space \citep[][]{TEGMARK2004,PERCIVAL2004}. 
In addition, a wrong assumption on the galaxy density contrast normalization will yield a non-vanishing mean resulting in erroneous large scale power when estimating power-spectra.
It is therefore generally crucial to account for these effects when inferring the density field and power spectrum from realistic galaxy surveys.

Traditional approaches generally measure the power spectrum and the galaxy bias parameters in separate steps by either measuring it from the same data set or employing fitting functions calibrated
on other galaxy catalogs \citep[see e.g. ][]{PERCIVAL2004,TEGMARK2004,COLE2005,PERCIVAL2007}. The risk of such approaches is that they may over-use the data
by ignoring joint and correlated uncertainties for inferred biases and power-spectra, leading to wrong conclusions on the accuracy of inferred quantities and possibly biased results because not all parameters are explored jointly. 

In principle there may also be cases where a joint inference of power spectrum and galaxy biases may be mutually beneficial. Traditional approaches would miss such opportunities.

In this paper we aim for the full characterization of the joint posterior \(\mathcal{P}\left(\{\delta_i\}, S ,\{\bar{N}^l\},\{b^l\}|\{N^l_i\}\right)\) \newtext{of \(M^{\mathrm{tracer}}\) different tracer populations labeled by the index \(l\)}. We will achieve this by  simultaneously exploring  
the three dimensional density field \(\delta_i\), the corresponding power spectrum \(S\), the
galaxy biases \(b^l\) and the galaxy density contrast normalizations \(\bar{N}^l\) given  the data -- galaxy observations in the form of three dimensional number counts \(N^l_i\).

\subsection{Dissecting the posterior distribution}
Direct analysis of the joint posterior  \(\mathcal{P}\left(\{\delta_i\}, S ,\{\bar{N}^l\},\{b^l\}|\{N^l_i\}\right)\) is challenging: it is non-Gaussian  and very high-dimensional. From a technical point of view, it is also not advisable to directly
generate random variates from the joint posterior distribution since this may result in a numerically unfeasible algorithm preventing rapid exploration of the parameter space.
In such a situation one usually has to rely on Metropolis-Hastings algorithms, which reduce the problem of numerical parameter space exploration 
to the design of suitable proposal distribution for the generation of candidate solutions. A particular important theorem on Metropolis Hastings block sampling permits to
dissect the high dimensional sampling problem into a sequence of lower dimensional problems once conditional distributions for the subtasks can be formulated \citep[][]{hastings}.
It is therefore possible to subdivide the exploration of the full joint parameter space into the following sequence of conditional sampling procedures:
\begin{eqnarray}
\label{eq:multiblock}
& {\rm{1}}) & \{\delta_i\}^{(s+1)}\curvearrowleft \mathcal{P}\left(\{\delta_i\}| S^s ,\{\bar{N}^l\}^s,\{b^l\}^s,\{N^l_i\}\right) \nonumber \\
& {\rm{2}}) & S^{(s+1)}\curvearrowleft \mathcal{P}\left(S^s| \{\delta_i\}^s ,\{\bar{N}^l\}^s,\{b^l\}^s,\{N^l_i\}\right) \nonumber \\ 
& {\rm{3}}) & \{\bar{N}^l\}^{(s+1)}\curvearrowleft \mathcal{P}\left(\{\bar{N}^l\}| S^s ,\{\delta_i\}^s,\{b^l\}^s,\{N^l_i\}\right) \nonumber \\
& {\rm{4}}) & \{b^l\}^{(s+1)}\curvearrowleft \mathcal{P}\left(\{b^l\}| S^s ,\{\delta_i\}^s,\{\bar{N}^l\}^s,\{N^l_i\}\right)\, ,
\end{eqnarray}
\newtext{where \(s\) labels the sample number.}
Iterating these individual sampling steps will then provide samples from the full joint posterior distribution \citep[][]{hastings}. Also see figure \ref{fig:flowchart}
where we have illustrated this multiple block sampling procedure, also clarifying the flow of information in subsequent iterations.
In the following we will discuss the individual contributions to the full joint posterior distribution as listed in equation (\ref{eq:multiblock}).

\subsection{The density posterior}
We now derive the posterior distribution to infer the three dimensional density field \citep[for further details see ][]{JASCHE2010PSPEC}.

\subsubsection{The density prior distribution}
This work mainly concerns the inference of the largest scales, where the density field can be reasonably well described by Gaussian statistics.
A Gaussian prior will therefore be an adequate description of the statistical behavior of the large scale density field.
It should nevertheless be remarked, that a Gaussian prior resembles the least informative prior, once the mean
and the covariance of the density field are specified. For this reason, from a Bayesian perspective, the Gaussian prior is well justified
approximation even for inference of the density field in the non-linear regime.
Therefore, in the following, we will assume the prior distribution for the density contrast \(\delta_i\) to be a multivariate normal distribution with zero mean and the Fourier transform of the covariance matrix \(S\)
being the cosmological power spectrum \(P(k)\). With these definitions the prior distribution for the density contrast can be written as: 
\begin{equation}
\label{eq:dens_prior}
\mathcal{P}\left(\{\delta_i\}|S\right)=\frac{\mathrm{e}^{-\frac{1}{2}\sum_{ij}^{N_{vox}} \delta_i\, S^{-1}_{ij} \, \delta_j}}{\sqrt{\mathrm{det}\left(2\,\pi\,S\right)}}\, .
\end{equation}
For further discussions the interested reader is referred to \citet{JASCHE2010PSPEC}.

\subsubsection{The density likelihood and the data model}
\label{datamodel_likelihood}
As already discussed above, it is crucial to account for different clustering behavior of galaxy populations in the large scale structure sample from which density fields and
power-spectra are inferred. In order to do so, we split the galaxy sample  into sub samples for which we can treat the respective systematic and statistical uncertainties.
The aim of this work is to present a method which can extract joint information on the three dimensional density contrast field \(\delta\), the power spectrum \(P(k)\), the bias \(b\) and the mean number
of galaxies in the survey \(\bar{N}\) from  a set of galaxy samples, all tracing the underlying density field differently, while properly accounting for the individual systematic and stochastic uncertainties.
We present the method in a general fashion that includes  joint cosmological analyses from two or more different galaxy surveys or even different probes of the large scale structure, while accounting for their respective systematic and stochastic uncertainties.
In the case of \(M^{\mathrm{tracer}}\) sub samples or galaxy observations, a linear data model for each of the subsets can be formulated as:
\begin{equation}
N^l_i=\bar{N}^l\,R^l_i\,\left(1+b^l\, D^{+}_i \,\delta_i \right) + \epsilon^l_i \, ,
\label{eq:data_model}
\end{equation}
where \(N^l_i\) are the observed number counts of galaxies in the \(l\)th sample and in the \(i\)th volume element, \(\bar{N}^l\) is the corresponding expected number of galaxies, \(R^l_i\) is the corresponding
survey response function, which accounts for survey geometry and selection function, \(b^l\) is a linear galaxy bias, \(D^{+}_i\) is the linear growth function and \(\epsilon^l_i\) is a random noise component. Also note, that here we are mainly interested
in recovering the largest scales and so a linear bias should suffice. We will defer the inclusion of more complicated non-linear and non-local biases to a future work.
The survey response operator \(R^l_i\) is given by the product of the angular mask \newtext{or completeness function} \(C^l_i\) and the radial selection function \(F^l_i\) at the \(i\)th volume element:
\begin{equation}
R^l_i=C^l_i\,F^l_i \, .
\end{equation}
By following the arguments in \citet{JASCHE2010PSPEC}, we will assume the random noise component \(\epsilon^l_i \) to follow a Gaussian distribution with zero mean and noise covariance matrix \(L^l_{ij}\), given as:
\begin{equation}
L^l_{ij}= \bar{N}^l\,R^l_i \,\delta^K_{ij} \, ,
\label{eq:cov_mat_noise}
\end{equation}
where \(\delta^K_{ij}\) is the Kronecker delta. Equation (\ref{eq:cov_mat_noise}) resembles the Covariance matrix of a Poisson process when averaged over realizations of the density contrast \(\delta_i\), since
the average of \(\delta_i\) vanishes \citep[see][for some discussions of the noise covariance matrix]{JASCHE2010PSPEC}. \newtext{Alternatively, the assumed Gaussian uncertainty model can be derived by a Taylor-expansion of the Poissonian log-likelihood to second order in the density field. This demonstrates, that the Gaussian likelihood model is correct at large scales, where density amplitudes are small, while it describes an approximation to small scale uncertainties. Although this Gaussian model assumes independence of the noise from the underlying density field, the number counts \(N^l_i\) of the \(M^{\mathrm{tracer}}\) tracer populations are correlated in spatially overlapping regions as they all refer to the same underlying three dimensional density field \(\delta_i\).
In this fashion, the data model given in equation (\ref{eq:data_model}) emulates the expected behavior of galaxy formation models. In this work, we are focusing at inferring the largest scales from observations, for which our data model is adequate. Note, that precise inference of the large scale structure
in the non-linear regime while accounting for the full Poissonian noise statistics can be performed with the method presented in \citet{JASCHE2010HADESMETHOD} and \citet{JASCHE2010HADESDATA}. }

With these definitions the likelihood for the \(l^{\mathrm{th}}\) dataset can be expressed as a multivariate normal distribution:
\begin{eqnarray}
\label{eq:density_likelihood}
\mathcal{P}^l\left(\{N^l_i\}|\{\delta_i\},\bar{N}^l,b^l\right) &=& \frac{\mathrm{e}^{-\frac{1}{2}\sum_{ij}^{N_{vox}} \left(N^l_i-\bar{N}^l\,R^l_i\,\left(1+b^l\,\delta_i \right)\right) L^{-1l}_{ij} \left(N^l_j-\bar{N}^l\,R^l_j\,\left(1+b^l\,\delta_j \right)\right)}}{\sqrt{\mathrm{det}\left(2\,\pi\, L\right)}} \nonumber \\
&=& \frac{\mathrm{e}^{-\frac{1}{2}\sum_{i}^{N_{vox}} \frac{\left(N^l_i-\bar{N}^l\,R^l_i\,\left(1+b^l\,\delta_i \right)\right)^2}{\bar{N}^l\,R^l_i}}}{ \left(2\,\pi\,\bar{N}^l\right)^{\frac{N_{vox}}{2}} \prod_i^{N_{vox}} \sqrt{R^l_i} } \nonumber \\
\label{eq:LH_per_dataset}
\end{eqnarray}     
The joint likelihood from all \(M^{\mathrm{tracer}}\) datasets is then obtained as:
\begin{eqnarray}
\mathcal{P}\left(\{N^l_i\}|\{\delta_i\},\{\bar{N}^l\},\{b^l\}\right) &=& \prod_l^{M^{\mathrm{tracer}}} \mathcal{P}^l\left(\{N^l_i\}|\{\delta_i\},\bar{N}^l,b^l\right)\,
\label{eq:LH_all_dataset}
\end{eqnarray}     
where we assumed conditional independence of the individual likelihoods, once the density contrast \(\delta_i\), the mean galaxy numbers \(\bar{N}^l\) and the biases \(b^l\) are given.

\subsection{The Power-spectrum posterior distribution}
As described above, sampling from the Wiener posterior distribution will provide realizations of full three dimensional density fields conditioned on the observations.
Basing on these density fields, in subsequent steps realizations of the corresponding power spectrum can be generated by sampling from
the conditional power spectrum posterior distribution:
\begin{eqnarray}
\label{eq:cond_ps_posterior}
\mathcal{P}\left(S| \{\delta_i\} ,\{\bar{N}^l\},\{b^l\},\{N^l_i\}\right) &=& \mathcal{P}\left(S\right) \mathcal{P}\left(\{\delta_i\}|S\right)  \frac{\mathcal{P}\left(\{\bar{N}^l\},\{b^l\},\{N^l_i\}| \{\delta_i\} \right)}{\mathcal{P}\left(\{\delta_i\} ,\{\bar{N}^l\},\{b^l\},\{N^l_i\}\right)}\nonumber \\ 
&\propto& \mathcal{P}\left(S\right) \mathcal{P}\left(\{\delta_i\}|S\right) \nonumber \\
&\propto& \mathcal{P}\left(S\right) \, \frac{\mathrm{e}^{-\frac{1}{2}\sum_{ij}^{N_{vox}} \delta_i\, S^{-1}_{ij} \, \delta_j}}{\sqrt{\mathrm{det}\left(2\,\pi\,S\right)}} \, ,
\end{eqnarray}
where we assumed conditional independence of the data on the power spectrum, once the full three dimensional density field is given, and in the last line we used the density prior given in equation (\ref{eq:dens_prior}).
The result provided by equation (\ref{eq:cond_ps_posterior}) demonstrates that once a realization of the three dimensional density field is given, inferring a realization of the power spectrum
is conditionally independent on the data. This fact was first pointed out by \citet{WANDELT2004} in case of CMB analysis and was later adapted for large scale structure inference by \citet{JASCHE2010PSPEC}. This is a particularly important result for the power spectrum sampling procedure. Since all observational systematics and uncertainties
have already been accounted for in inferring a density field realization, this considerably simplifies conditional power spectrum inference \citep[also see ][ for a deeper discussion]{JASCHE2010PSPEC}.
Also note, that equation (\ref{eq:cond_ps_posterior}) is a standard result for Bayesian density and power spectrum inference with Wiener posteriors in the case of CMB and large scale structure analyses \citep[see e.g.][]{WANDELT2004,2004ApJS..155..227E,JASCHE2010PSPEC}.
It is interesting to remark that equation (\ref{eq:cond_ps_posterior}) provides a direct sampling procedure to generate power spectrum realizations.
Following \citet{JASCHE2010PSPEC}, equation (\ref{eq:cond_ps_posterior}) can be re expressed as an inverse Gamma distribution, when employing a power-law prior for the power spectrum amplitudes.
Thus, inserting a power law prior in equation (\ref{eq:cond_ps_posterior}) and imposing the correct normalization, the conditional power spectrum posterior can be written as \citep{JASCHE2010PSPEC}:
\begin{equation}
\label{eq:INVERSE_GAMMA}
{\cal P}(P_m|\{s_i\}) =\frac{\left(\frac{\sigma_m}{2}\right)^{(\alpha-1)+n_m/2}}{\Gamma \left((\alpha-1)+\frac{n_m}{2}\right)} \frac{1}{P_m^{(\alpha+n_m/2)}} e^{- \frac{1}{2}\frac{\sigma_m}{P_m}}\, ,
\end{equation}
where \(P_m\) is the power spectrum amplitude of the \(m\)th bin corresponding to the Fourier mode \(k_m\), \(n_m\) is the number of modes in that bin, \(\alpha\) is the power-law index of the power-law prior, \(\Gamma(x)\) is the Gamma function
and \(\sigma_m\) is given as \newtext{the squared isotropic average over the m{\it{th}} mode bin in Fourier space } :
\begin{equation}
\sigma_m=\sigma(|\vec{k}_m|)=\sum_{\phi} \sum_{\theta} \left | \hat{\delta}(|\vec{k}_m|, \phi, \theta) \right |^2 \, ,
\end{equation}
with \(\hat{\delta}(\vec{k})\) being the Fourier transform of the density field. \newtext{Note, that the power-law prior with \(\alpha=0\) describes a "flat" prior on a unit scale, while \(\alpha=1\) yields a Jeffreys Prior. Jeffreys prior is a solution to a measure invariant scale transformation, and hence is a scale independent prior, as different scales have the same probability.
For this reason, Jeffreys prior is optimal to infer cosmological power-spectra, which constitute scale measurements, as it does not introduce any bias on a logarithmic scale.
}
For a more detailed derivation and discussion of the Inverse Gamma distribution in case of LSS inference the reader is referred to our previous work presented in \citet{JASCHE2010PSPEC}.   

\begin{figure*}
\centering{\includegraphics[width=1.0\textwidth,clip=true]{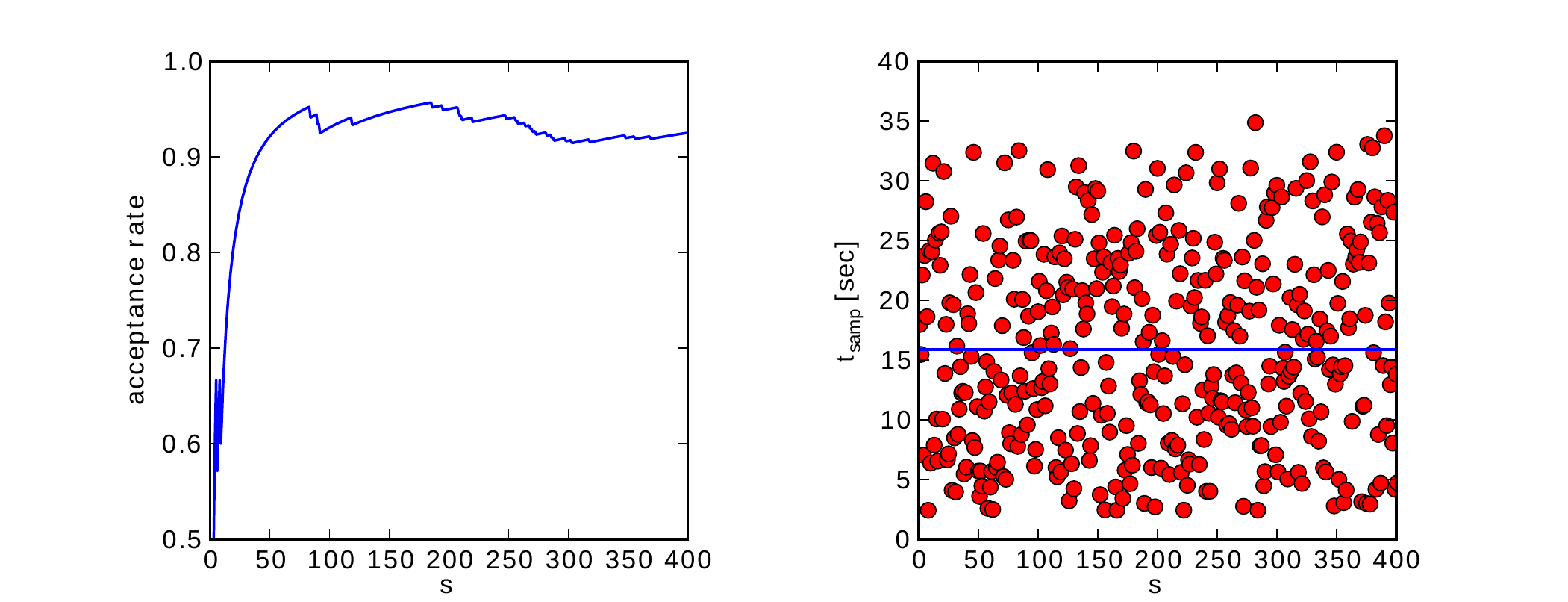}}
\caption{Acceptance rate for successive samples (left panel) and the execution time per sample (right panel). It can be seen that the acceptance rate \newtext{increases} during the initial burn in phase and finally stabilizes at about \newtext{\(92\)} per cent.
The left panel demonstrates the scatter in the execution times of individual samples. The average execution time is about \newtext{16} seconds as indicated by the solid blue line.}
\label{fig:accept_rat}
\end{figure*}

\subsection{The bias posterior distribution}
\label{sample_the_bias}
Accounting for galaxy biases is particularly important when analyzing galaxy samples exhibiting widely  varying clustering behavior across galaxy types.
These galaxy biases give rise to additional systematics, which when ignored, have the potential to distort the shape of inferred power-spectra.
We extend the power spectrum sampling framework described in \citet{JASCHE2010PSPEC} with a galaxy bias sampling procedure.
To be maximally agnostic about the  functional shape of the galaxy bias as a function of galaxy properties such as luminosity or color, we will assign a 
galaxy bias \(b^l\) to each of the \(M^{\mathrm{tracer}}\) individual galaxy sub samples. Another approach would be to assume  a functional shape for the galaxy bias as
a function of some galaxy properties such as luminosity and just fit the corresponding function parameters. Such an approach would be \textit{a posteriori} and hence overuse the data since knowledge of such a functional shape
is generally extracted from the galaxy sample under consideration itself.
We will instead pursue an approach that only assumes that bias properties remain approximately constant within luminosity bins. 

Realizations of the bias factors \(b_l\) can then be obtained by sampling from the conditional bias posterior distribution given as:
\begin{equation}
\mathcal{P}\left(\{b^l\}| S ,\{\delta_i\},\{\bar{N}^l\},\{N^l_i\}\right) = \mathcal{P}\left(\{b^l\}|\{\delta_i\},\{\bar{N}^l\},\{N^l_i\}\right) \, ,
\end{equation}
where the bias factors \(b_l\) are conditionally independent of the power spectrum, once the full three dimensional density field is known.
Using Bayes theorem we can rewrite the conditional bias posterior as:
\begin{eqnarray}
\label{eq:bias_posterior_joint}
\mathcal{P}\left(\{b^l\}|\{\delta_i\},\bar{N}^l,\{N^l_i\}\right) &=& \mathcal{P}\left(\{b^l\}\right)\, \mathcal{P}\left(\{\delta_i\},\{\bar{N}^l\}|\{b^l\}\right) \nonumber \\
& & \times \frac{\mathcal{P}\left(\{N^l_i\}|\{\delta_i\},\{\bar{N}^l\},\{b^l\}\right)}{\mathcal{P}\left(\{\delta_i\},\{\bar{N}^l\},\{N^l_i\}\right)} \nonumber \\  
&\propto & \prod_l \mathcal{P}\left(b^l\right)\, \mathcal{P}\left(\{N^l_i\}|\{\delta_i\},\bar{N}^l,b^l\right) \, ,
\end{eqnarray}  
where we introduced a prior factorizing in the different galaxy samples \(\mathcal{P}\left(\{b^l\}\right)=\prod_l \mathcal{P}\left(b^l\right)\) and assumed
conditional independence \(\mathcal{P}\left(\{\delta_i\},\{\bar{N}^l\}|\{b^l\}\right)=\mathcal{P}\left(\{\delta_i\},\{\bar{N}^l\}\right)\).
Since the probability distribution given in equation (\ref{eq:bias_posterior_joint}) factorizes in the \(M^{\mathrm{tracer}}\) galaxy sub samples, one can estimate the bias
factors for each galaxy sample separately by sampling from the following distribution:
\begin{eqnarray}
\mathcal{P}\left(b^l|\{\delta_i\},\bar{N}^l,\{N^l_i\}\right) &\propto& \mathcal{P}\left(b^l\right)\, \mathcal{P}\left(\{N^l_i\}|\{\delta_i\},\bar{N}^l,b^l\right) \, .  
\end{eqnarray}
In order to follow the maximum ignorance principle also with regard to the amplitude of the bias, we assume a flat prior \(\mathcal{P}\left(b^l\right)\propto\,1\).
With these assumptions the conditional bias posterior distribution is simply proportional to the density likelihood given in equation (\ref{eq:density_likelihood}):
\begin{eqnarray}
\label{eq:galaxybias_post}
\mathcal{P}\left(b^l|\{\delta_i\},\bar{N}^l,\{N^l_i\}\right) &\propto& \mathrm{e}^{-\frac{1}{2}\sum_{i}^{N_{vox}} \frac{\left(N^l_i-\bar{N}^l\,R^l_i\,\left(1+b^l\,\delta_i \right)\right)^2}{\bar{N}^l\,R^l_i}} \nonumber \\
\end{eqnarray}  
As can  be seen from equation (\ref{eq:galaxybias_post}) and as demonstrated in appendix \ref{bias_gauss}, this distribution turns into a univariate normal distribution for the bias factor \(b^l\) given as: 
\begin{eqnarray}
\mathcal{P}\left(b^l|\{\delta_i\},\bar{N}^l,\{N^l_i\}\right) = \frac{\mathrm{e}^{-\frac{1}{2} \frac{ \left( b^l - \mu_{b^l} \right)^2 }{\left(\sigma_{b^l}\right)^2}}}{\sqrt{2\,\pi \left(\sigma_{b^l}\right)^2} } \,,\nonumber \\
\end{eqnarray}
with \(\sigma_{b^l}\) given as:
\begin{equation}
\left(\sigma_{b^l}\right)^2 = \frac{1}{\sum_{i}^{N_{vox}} \bar{N}^l\,R^l_i\,\delta_i} \, , 
\end{equation}
and the mean \( \mu_{b^l}\)  given as:
\begin{equation}
\mu_{b^l} =\frac{\sum_{i}^{N_{vox}} \left(N^l_i-\bar{N}^l\right)\,\delta_i}{\sum_{i}^{N_{vox}} \bar{N}^l\,R^l_i\,\delta_i} \, .
\end{equation}
A random realization for the galaxy bias can thus be obtained by simply sampling from a univariate normal distribution.
Note, that while previous methods such as presented in \citet{TEGMARK2004} measure the relative bias from estimated power-spectra, our method
infers the bias parameters directly from the relation between the data and the three dimensional density field. Further, since the density field
contains information inferred jointly from all \(M^{\mathrm{tracer}}\) galaxy sub samples, inference of individual bias factors depends on all other biases via the density field, enabling us  to measure relative biases and thus  the shape of the power spectrum up to an overall normalization.

\begin{figure*}
\centering{\includegraphics[width=0.8\textwidth,clip=true]{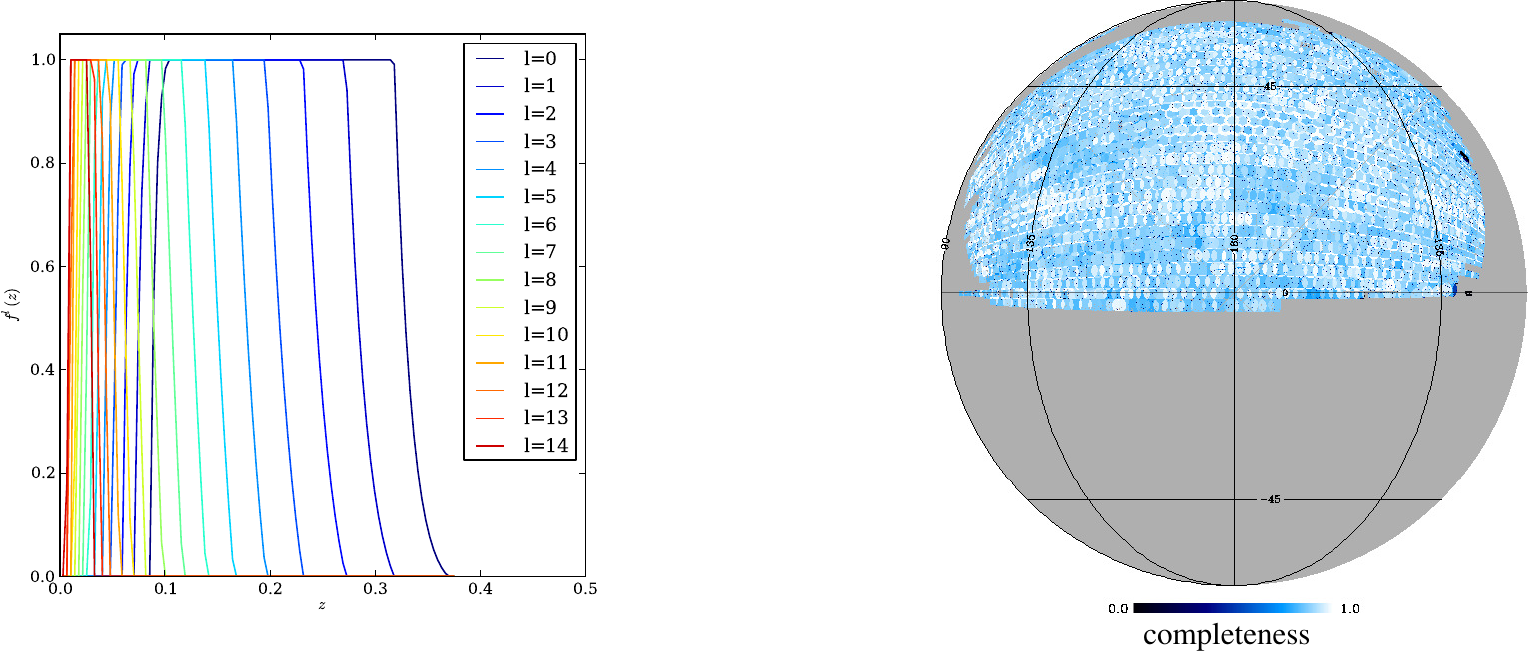}}
\caption{Selection functions for 15 absolute magnitude bins (left panel) and the SDSS survey geometry (right panel).}
\label{fig:selfunc_surveygeometry}
\end{figure*}

\subsection{The \(\bar{N}\) posterior distribution}
\label{sample_nmean}
Of similar importance as galaxy biases are the mean numbers of galaxies \(\bar{N}^l\) for the \(M^{\mathrm{tracer}}\) sub samples.
These mean numbers of galaxies \(\bar{N}^l\) are crucial to define the true underlying density contrast of the galaxy distribution. A false value of \(\bar{N}^l\) will yield a non-vanishing mean and result in 
erroneous large scale power in the inferred power-spectra. 
Since the \(\bar{N}^l\) are not known a priorly, they will also have to be inferred jointly, in order not to miss possible cross-correlations and interdependencies.
The parameters \(\bar{N}^l\) for the \(M^{\mathrm{tracer}}\) sub samples can then be inferred from a conditional distribution given as:
\begin{equation}
\mathcal{P}\left(\{\bar{N}^l\}| S ,\{\delta_i\},\{b^l\},\{N^l_i\}\right)=\mathcal{P}\left(\{\bar{N}^l\}|\{\delta_i\},\{b^l\},\{N^l_i\}\right) \, , \nonumber \\
\end{equation}
where, in a similar fashion as described in section \ref{sample_the_bias}, we assume conditional independence of the power spectrum once the full three dimensional density field is given.
Further, following similar arguments as described in section \ref{sample_the_bias}, it is possible to infer each parameter \( \bar{N}^l \) separately, since the corresponding 
probability distribution factorizes. As a result we obtain the following probability distribution for the parameter \( \bar{N}^l \):
\begin{eqnarray}
\mathcal{P}\left(\bar{N}^l|\{N^l_i\},\{\delta_i\},b^l\right)&\propto& \mathcal{P}\left(\bar{N}^l\right)\,\mathcal{P}\left(\{\{N^l_i\}|\delta_i\},b^l,\bar{N}^l\right) \nonumber \\
&\propto & \frac{\mathrm{e}^{-\frac{1}{2}\sum_{i}^{N_{vox}} \frac{\left(N^l_i-\bar{N}^l\,R^l_i\,\left(1+b^l\,\delta_i \right)\right)^2}{\bar{N}^l\,R^l_i}}}{\left(\bar{N}^l\right)^{\frac{N_{vox}}{2}} } \, . \nonumber \\
\label{eq:nmean_post}
\end{eqnarray}
Note, that also here we followed a maximum ignorance approach by setting the prior \(\mathcal{P}\left(\bar{N}^l\right)\propto1\). As can be seen the resultant 
probability distribution for the parameter \(\bar{N}^l\) is proportional to the density likelihood given in equation (\ref{eq:LH_per_dataset}).
Following the discussion presented in appendix \ref{inv_gauss}, the probability distribution for the \(\bar{N}^l\) given in equation (\ref{eq:nmean_post}) 
is a generalized inverse Gaussian distribution (GIG) given as:
\begin{eqnarray}
\mathcal{P}\left(\bar{N}^l|\{N^l_i\},\{\delta_i\},b^l\right)= \left(\frac{A^l}{B^l}\right)^{\frac{p^l}{2}} \frac{\left(\bar{N}^l\right)^{p^l-1}}{2K_{p^l}\left(\sqrt{A^l\,B^l}\right)}\,\mathrm{e}^{-\frac{1}{2} \left( \bar{N}^l A^l+ \frac{1}{\bar{N}^l} B^l  \right)} \, \,
\label{eq:GIG_post}
\end{eqnarray}
where \(K_{p^l}(x)\) is a modified Bessel function of the second kind, \(p^l=1-N^{obs}_{vox}/2\), \(A^l\) is given as:
\begin{equation}
A^l=\sum_{i}^{N_{vox}} R^l_i\,\left(1+b^l\,\delta_i \right)^2 \, ,
\end{equation}
and \(B^l\) being:
\begin{equation}
B^l=\sum_{i}^{N_{vox}} \left(N^l_i\right)^2/R^l_i\, .
\end{equation}
\newtext{The number of observed voxels \(N^{obs}_{vox}\) is given by:}
\begin{equation}
N^{obs}_{vox} = \sum_i^{N_{vox}} \theta(R^l_i) \, ,
\end{equation}
\newtext{where the Heaviside function \(\theta(x)\) counts the non zero elements of the three dimensional survey mask.}
Random realizations for the parameters \(\bar{N}^l\) can then be provided by sampling from the GIG distribution.
It is interesting to note, that the inferred \(\bar{N}^l\) also depend on the galaxy bias parameters and
the three dimensional density field. This indicates, that a joint estimation of all these parameters is required
in order to correctly account for joint information and correlated uncertainties. 
Algorithms to sample from the GIG distribution are described in literature \citep[see e.g.][]{DAGPUNAR1988}.

\begin{figure*}
\centering{\includegraphics[width=1.1\textwidth,clip=true]{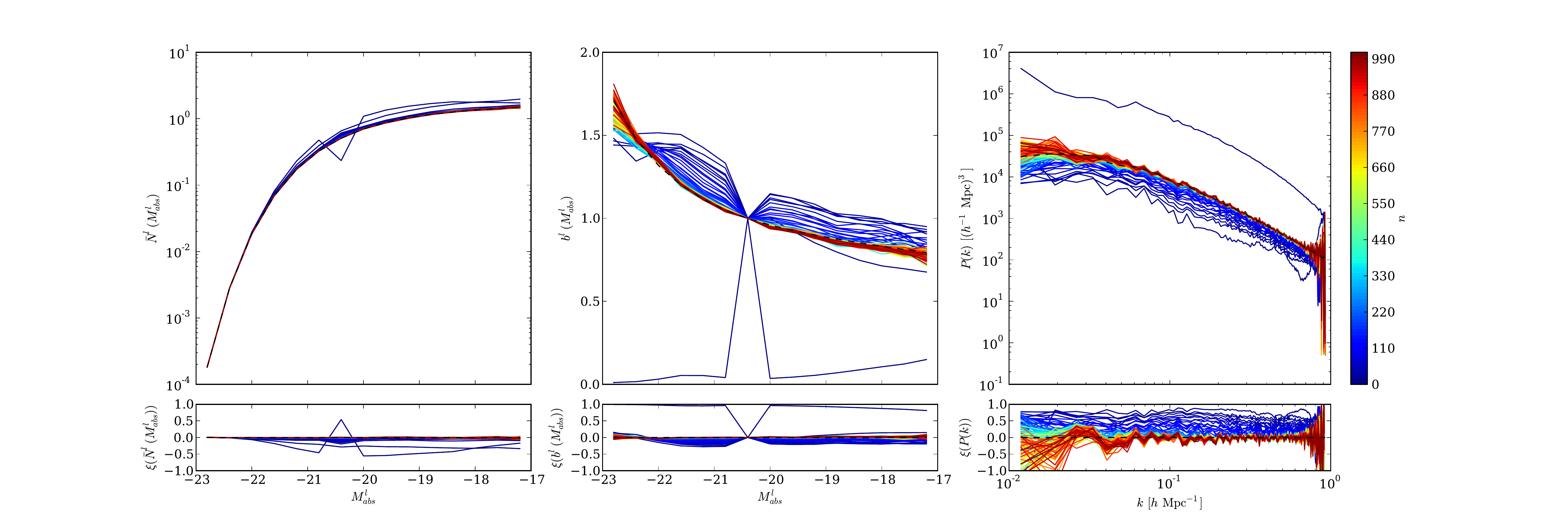}}
\caption{The plots demonstrate the burn-in behavior of the algorithm by following the systematic drift of successive \(\bar{N}^l\) (left panels), \(b^l\) (middle panels) and \(P(k)\) (right panels) samples during the burn-in phase. Successive samples are color coded corresponding to their sample number as indicated by the color bar on the right. Black dashed lines correspond to the respective true underlying values.
Lower panels depict the successive deviation \(\xi\) from the true values, as described in the text, for the  \(\bar{N}^l\), \(b^l\) and \(P(k)\) samples respectively.
The sequence of 1000 successive samples, visualizes how the sampler approaches the true underlying values and starts exploring the parameter space around them, corresponding to the respective uncertainties.}
\label{fig:burnin}
\end{figure*}

\section{Numerical implementation}
\label{numericalimplementation}
In this section we will describe the numerical implementation of the large scale structure sampler.

\subsection{Sampling the density field}
\label{sample_dens}
As is manifest from the above discussion, the three dimensional density sampling step resembles the core of our method. The Wiener posterior for the density sampling
step described in section \ref{LSS_POSTERIOR} is a multivariate normal distribution, and as such possesses a clear sampling procedure, by first 
calculating the Wiener-mean and adding a variance term corresponding to the Wiener-variance \citep[see e.g.][]{WANDELT2004,ODWYER2004,2004ApJS..155..227E,JEWELL2004,LARSON2007,ERIKSEN2007,JEWELL2009,JASCHE2010PSPEC}. 
Nevertheless, calculating the Wiener mean and variance requires matrix inversions for large matrices with typically on the order of \( 10^7 \times 10^7\) entries.
These matrix inversions can be performed with efficient operator based implementations of conjugate gradient methods \citep[see e.g.][]{WANDELT2004,2004ApJS..155..227E,KITAURA2009,JASCHE2010PSPEC}.
Alternatively, the Wiener posterior in case of CMB data has also been explored using a Hybrid Monte Carlo (HMC) algorithm \citep[][]{TAYLOR2008}.
Furthermore, the HMC has been successfully applied for the inference of three dimensional density fields in the non-linear regime \citep[see e.g.][]{JASCHE2010HADESMETHOD,JASCHE2010HADESDATA,JASCHE2012}.
Since these references have found the HMC algorithm to be very efficient in  large scale structure problems we implement an HMC algorithm for the exploration of the large scale structure density field.
Compared to the direct sampling approach based on  iterative techniques,   HMC is less sensitive to the effects of numerical inaccuracies, e.g. due to insufficiently strict tolerances in the inversion scheme, because the final accept/reject step of the Metropolis Hastings sampler ensures correctness of the target density. Any such inaccuracies would only impact efficiency but not correctness. 

In the following we will  review the basics of the  HMC algorithm required for this work.

\subsubsection{The HMC algorithm}
\label{HAMILTONIAN_SAMPLING}
The numerical efficiency of the Hybrid Monte Carlo algorithm originates from the fact that it exploits techniques developed to follow classical dynamical particle motion in potentials to deterministically
provide new proposals to the Metropolis-Hastings algorithm \citep[][]{DUANE1987,NEAL1993,NEAL1996}.
Assuming we wish to generate random variates from a  probability distribution \({\mathcal P}(\{x_i\})\), with \(\{x_i\}\) being a set consisting of the \(N\) elements \(x_i\), we may
interpret the negative logarithm of this distribution as a potential \(\psi(x)\):
\begin{equation}
\label{eq:Potential}
\psi(x)=-ln({\mathcal P}(x)) \, ,
\end{equation}
Further, by introducing a 'momentum' variable \(p_i\) and a 'mass matrix' \(M\), one can  add a kinetic term to the potential \(\psi(x)\) in order to obtain a Hamiltonian function, which, similar to ordinary mechanics, describes the dynamics in a high dimensional parameter space.
Such a Hamiltonian can be written as:
\begin{equation}
\label{eq:Hamiltonian}
H = \sum_i\sum_j \frac{1}{2}\,p_i\,M_{ij}^{-1}\,p_j +\psi(x) \, .
\end{equation}
The new target probability distribution is then obtained by exponentiating the Hamiltonian given in equation (\ref{eq:Hamiltonian}):
\begin{equation}
\label{eq:TARGET_DISTRIBUTION}
e^{-H} = {\mathcal P}(\{x_i\})\,e^{-\frac{1}{2}\,\sum_i\sum_j\,p_i\,M_{ij}^{-1}\,p_j}\, .
\end{equation}
As can be seen from equation (\ref{eq:TARGET_DISTRIBUTION}), the form of the Hamiltonian was chosen such, that the new target distribution is separable into a Gaussian distribution in the momenta \(\{p_i\}\) and the target distribution \({\mathcal P}(\{x_i\})\).
This immediately clarifies that the two sets of variables \(\{p_i\}\) and \(\{x_i\}\) are independent and marginalization over the artificial momenta yields the original target distribution \({\mathcal P}(\{x_i\})\).

Given an initial point in this high dimensional phase space \( (\{x_i\},\{p_i\})\) it is possible to deterministically propose a new sample
from the target distribution by following persistent trajectories for some fixed time \(\tau\) according to Hamilton's equations:
\begin{equation}
\label{eq:HAMILTON1}
\frac{dx_i}{dt} = \frac{\partial H}{\partial p_i}\, ,
\end{equation}
and
\begin{equation}
\label{eq:HAMILTON2}
\frac{dp_i}{dt} = \frac{\partial H}{\partial x_i} = - \frac{\partial \psi(x)}{\partial x_i}\, .
\end{equation}
A new point in phase space will be accepted according to the standard Metropolis-Hastings acceptance rule:
\begin{equation}
\label{eq:acceptance_rule}
{\mathcal P}_A = min\left[1,exp(-\left(H(\{x'_i\},\{p'_i\})-H(\{x_i\},\{p_i\})\right)\right]\, .
\end{equation}
Since Hamiltonian dynamics conserve the Hamiltonian, the HMC approach results in an acceptance rate of unity. In practical applications, however,
the acceptance rate may be lower due to numerical inaccuracies in the integration scheme.
Once a new sample has been accepted the momentum variable is discarded and the process restarts by randomly drawing a new set of momenta.
The individual momenta \(\{p_i\}\) will not be stored, and therefore discarding them amounts to marginalizing over this auxiliary quantity.
Hence, the Hamiltonian sampling procedure basically consists of two steps. The first step is a Gibbs sampling step, which yields a new set of Gaussian distributed momenta. The second step, on the other hand amounts to solving a deterministic dynamical trajectory on the posterior surface.

\begin{figure*}
\centering{\includegraphics[width=1.0\textwidth,clip=true]{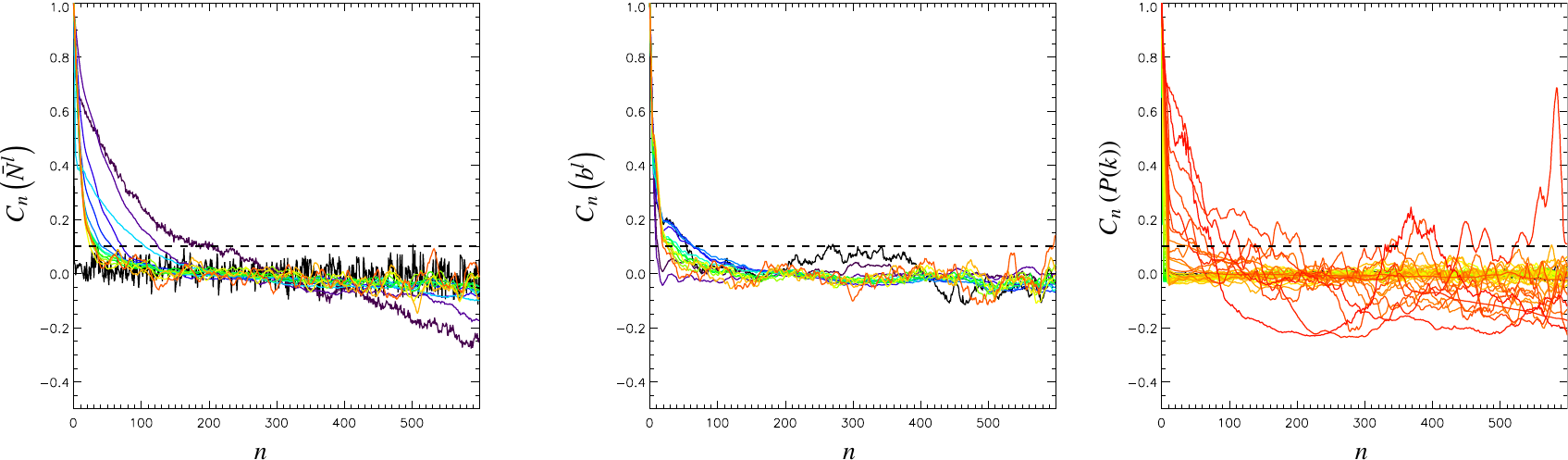}}
\caption{Correlation length for \(\bar{N}^l\) (left panel), \(\bar{b}^l\) (middle panel) and the power spectrum \(P(k)\) (right panel). Different line colors correspond to 15 different \(\bar{N}^l\) and \(\bar{b}^l\) parameter in the left and middle panel  and to 128 power-spectrum modes \(P(k)\) in the right panel.  It can be seen
that with typical correlation length of on the order of 100 samples the overall statistical efficiency of the algorithm is high.}
\label{fig:corrlength}
\end{figure*}

\subsubsection{Equations of motion for the Wiener Posterior}
\label{equations_of_motion}
As described above, in this work we will employ a Wiener posterior for the inference of the three dimensional density field.
In this case the Hamiltonian potential can be written as:
\begin{eqnarray}
\label{eq:Wiener_Posterior_psi}
\psi(\{\delta_i\})= \frac{1}{2} \sum_{ij}^{N_{vox}} \delta_i\, S^{-1}_{ij} \, \delta_j + \frac{1}{2} \sum_{i}^{N_{vox}} \sum_{l} \frac{\left(N^l_i-\bar{N}^l\,R^l_i\,\left(1+b^l\,\delta_i \right)\right)^2}{\bar{N}^l\,R^l_i} \, , \nonumber \\
\end{eqnarray}
where we omitted terms corresponding to normalization constants.
Variation of \(\psi(\{\delta_i\})\) with respect to the three dimensional density field \(\delta_p\) then yields the Hamiltonian forces which govern the
dynamics in parameter space:
\begin{eqnarray}
\label{eq:Wiener_Posterior_force}
\frac{\partial \psi(\{\delta_i\})}{\partial \delta_p} &=&  \sum_{j} \left[S^{-1}_{pj}+\sum_l \delta^K_{pj} (b^l)^2 \bar{N}^l\,R^l_j \right]\,\delta_j \nonumber \\
& & - \sum_{l} b^l\,\left( N^l_p-\bar{N}^l\,R^l_p \right)\, . \nonumber \\ 
\end{eqnarray}
With this definition the equations of motion can be given as:
\begin{equation}
\label{eq:HAMILTON1_a}
\frac{d\delta_i}{dt} = \sum_j M_{ij}^{-1} \,p_j\, ,
\end{equation}
and
\begin{eqnarray}
\label{eq:HAMILTON2_a}
\frac{dp_i}{dt} &=&  -\sum_{j} \left[S^{-1}_{pj}+\sum_l \delta^K_{pj} (b^l)^2 \bar{N}^l\,R^l_j \right]\,\delta_j \nonumber \\
& & + \sum_{l} b^l\,\left( N^l_p-\bar{N}^l\,R^l_p \right)\, . \nonumber \\
\end{eqnarray}
As can be seen from equation (\ref{eq:HAMILTON2_a}), in the framework of the HMC the quadratic form of the Wiener posterior yields 
a multidimensional coupled harmonic oscillator whose zero points are defined by the data.
Also note, that equations (\ref{eq:HAMILTON1_a}) and (\ref{eq:HAMILTON2_a}) can be combined to yield a second order differential equation:
\begin{eqnarray}
\label{eq:OSC_EQUATION}
\frac{d^2 \delta_h}{dt^2} &=& - \sum_{p} M^{-1}_{hp} \sum_{j} \left[S^{-1}_{pj}+\sum_l \delta^K_{pj} (b^l)^2 \bar{N}^l\,R^l_j \right]\,\delta_j \nonumber \\
& & + \sum_{p} M^{-1}_{hp} \sum_{l} b^l\,\left( N^l_p-\bar{N}^l\,R^l_p \right) \, . 
\end{eqnarray}
This clarifies that an ideal choice for the mass matrix \(M\) would be given by:
\begin{eqnarray}
\label{eq:Mass_matrix}
M_{ij} = S^{-1}_{ij}+\sum_l \delta^K_{ij} (b^l)^2 \bar{N}^l\,R^l_j \, ,
\end{eqnarray}
which would decouple the harmonic oscillators and permit a trivial analytic solution to the equations of motion for a set of independent harmonic oscillators.
Nevertheless, this approach would again require matrix inversions and hence be identical to a Gibbs sampling approach, which we try to avoid here \citep[see e.g.][]{WANDELT2004,ODWYER2004,2004ApJS..155..227E,JEWELL2004,LARSON2007,ERIKSEN2007,JEWELL2009,JASCHE2010PSPEC}.
In order to provide samples from the target distribution, we will therefore integrate the equations of motion (\ref{eq:HAMILTON1_a}) and (\ref{eq:HAMILTON2_a})
numerically \newtext{over a pseudo time interval \(\tau\)} via a leapfrog scheme. \newtext{In order to avoid resonant trajectories, the pseudo time interval \(\tau\) is drawn from a uniform distribution.} It is \newtext{also} important to remark, that the leapfrog scheme is a symplectic integrator, and as such ensures detailed balance of the Markov chain.
For a detailed description of the numerical implementation of the HMC in general and in cosmology in particular the reader is referred to the literature
\citep[in particular see][]{DUANE1987,NEAL1993,NEAL1996,TAYLOR2008,JASCHE2010HADESMETHOD}. 
In general the choice of the mass matrix does not affect the validity but only the numerical efficiency of the sampling method. We choose a diagonal form consisting of
the diagonal elements of the full mass matrix, given in equation (\ref{eq:Mass_matrix}), in its Fourier representation. This choice enhances performance while avoiding matrix inversions, analogous to using a diagonal preconditioner  in iterative conjugate gradients solvers.

\begin{figure*}
\centering{\includegraphics[width=0.5\textwidth,clip=true]{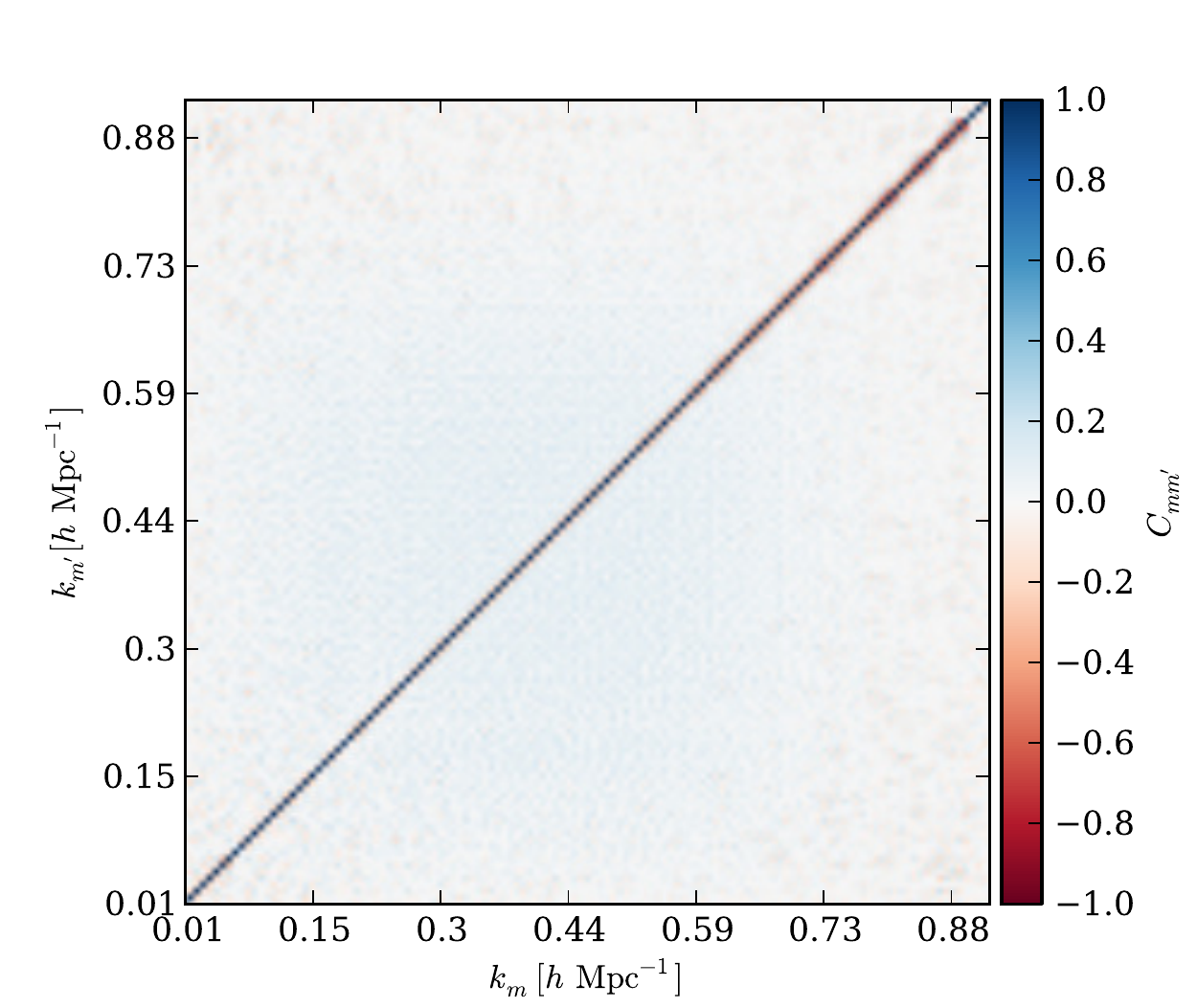}}
\caption{The plot shows the cross-correlation matrix \(\mat{C}_{l\,l'}\) between the different modes \(k_l\) of the power spectrum. As can be seen,
the matrix is strongly peaked near the diagonal with anti-correlation of adjacent spectral bins. These results demonstrate
the ability of our method to accurately determine the posterior correlation matrix for the power-spectrum amplitudes \(P(k)\) . }
\label{fig:corrmat}
\end{figure*}

\subsection{Sample the power spectrum}
\label{samplepowspec}
The conditional power spectrum posterior, given by the inverse gamma distribution in equation (\ref{eq:INVERSE_GAMMA}) , facilitates a direct sampling scheme  \citep[see e.g.][]{2004ApJS..155..227E,WANDELT2004,JASCHE2010PSPEC}.
In particular, by performing a change of variables \(x_m=\sigma_m/P_m\) one obtains the \(\chi^2\)-distribution\begin{equation}
\label{eq:XI_SQUARE}
{\cal P}(x_m|\{s_i\}) = \frac{x_m^{\beta_m/2-1}}{\Gamma\left(\beta_m/2\right)\,(2)^{\beta_m/2}} e^{- \frac{x_m}{2}}\, ,
\end{equation}
where \(\beta_m=2(\alpha+n_m/2-1)\).
Consequently, generating random variates for the power spectrum reduces to the task of drawing \newtext{independent} random samples from the \(\chi^2\)-distribution \newtext{for each Fourier mode \(k_m\)}.
Let  \(z_q\) be \(\beta_m\)  independent, normally distributed random variates with zero mean and unit variance. Then \citep[][]{JASCHE2010PSPEC}:
\begin{equation}
\label{post:posterior_of_x_1}
x_m = \sum_{q=1}^{\beta_m} z_q^2=|\vec{z}_m|^2 \,
\end{equation}
is \(\chi^2\)-distributed, with \(\vec{z}_m\)  a \(\beta_m\) element vector with elements $ z_q$.
A sample of the \newtext{power spectrum amplitude} \(P_m\) \newtext{ at Fourier mode \(k_m\)} is then obtained by:
\begin{equation}
\label{eq:PS_sample}
P_m = \frac{\sigma_m}{|\vec{z}_m|^2} \, .
\end{equation}
It should be remarked that each \(P_m\) is a positive quantity, ensuring the requirement that the signal covariance matrix be a  positive definite quantity.

\subsection{Exploring the low signal to noise regime}
In a Poissonian picture of galaxy formation the signal-to-noise ratio can be expressed as the square root of the number of galaxies \newtext{\(N_{gal}\)} found
in a given volume \(S/N=\sqrt{N_{gal}}\). Consequently, splitting a galaxy sample into a number of sub samples, to accurately account for the respective systematics
will yield a lower signal-to-noise ratio for each of these galaxy sub samples. Although, the Likelihood given in equation (\ref{eq:LH_all_dataset}) correctly 
accounts for the joint recovery of information from all galaxy samples by correctly weighting them corresponding to their respective 
noise properties, the overall signal-to-noise ratio may still be low, depending on the amount of sub samples.
This effect may influence the numerical efficiency for the power spectrum sampling procedure as described above. 

While the approach described in sections \ref{sample_dens} and \ref{samplepowspec} provably converges to the
target posterior on all scales probed by the data, it may take prohibitive computational time to generate a sufficient amount of independent samples in low signal to noise regime \citep{JEWELL2009}. 
In particular, the variations in subsequent power spectrum samples are solely determined by cosmic variance, whereas the full posterior distribution is 
governed by cosmic variance and noise. This permits a rapid exploration of parameters at the largest scales, where cosmic variance is dominant, but yields poor statistical efficiency for the small scale regime, where cosmic variance is typically much smaller than the noise variance. This results in a prohibitively long correlation length of the sequence of spectra in the low-signal to noise regime, requiring unfeasible long Markov Chains
to generate sufficient numbers of independent samples \citep{JEWELL2009}.
The above discussion shows that the variables chosen to explore the parameter space are optimal for the cosmic variance dominated regime,
but it may also suggest that a change of variables is advantageous for the exploration of the low signal to noise regime.

In fact, a change of variables can greatly improve the mixing behavior of the Markov Chain. Following \citet{JEWELL2009},
we introduce a change of variables which corresponds to the following  deterministic transition for the three dimensional density field in Fourier space:
\begin{equation}
\label{eq:det_trans}
\hat{\delta}^{s+1}_k = \sqrt{\frac{P^{s+1}(k)}{P^{s}(k)}} \hat{\delta}^{s}_k \, ,
\end{equation}
while, as demonstrated in appendix \ref{mixingefficiency}, a new power spectrum sample has to be generated from the distribution:  
\begin{eqnarray}
\label{eq:mixing_pdf}
\mathcal{P} \left(\{ P^{s+1}(k) \} | \{u_k\} , \{ d \} \right) & = & \mathcal{P}\left(\{ P^{s+1}(k)\} \right)  \nonumber \\
&  & \times  \prod_l \mathrm{e}^{-\frac{1}{2}\sum_{i}^{N_{vox}} \frac{\left(N^l_i-\bar{N}^l\,R^l_i\,\left(1+b^l\,C\sum \sqrt{P^{s+1}(k)} \, \hat{u}_k e^{2\pi i k \frac{\sqrt{-1}}{N}}  \right)\right)^2}{\bar{N}^l\,R^l_i}} \nonumber \\
\end{eqnarray}
with \(\hat{u}_k= \hat{\delta}^{s}_k / P^{s}(k) \) \newtext{and \(s\) labeling the sample number in the Markov chain}. Unfortunately there exists no immediate way to generate power spectrum samples from this distribution.
While \citet{JEWELL2009} propose a simple Metropolis-Hastings sampler to generate samples from the distribution given in equation (\ref{eq:mixing_pdf}), we will
employ a HMC to solve the problem. As already discussed above, the HMC has the advantage that it greatly improves the acceptance probability compared
to standard implementations of a Metropolis-Hastings algorithm. Furthermore, the entire procedure of sampling a power spectrum from the distribution given in equation (\ref{eq:mixing_pdf})
followed by the deterministic transition, described by equation (\ref{eq:det_trans}), for the density field is numerically as expensive as the sampling procedure outlined in sections \ref{sample_dens} and \ref{samplepowspec}.
Alternating these two sampling procedures will thus not decrease the overall \textit{numerical} efficiency of the algorithm. On the other hand, alternated sampling in
these two variable bases greatly improves the \textit{statistical} efficiency  and hence the overall performance of our algorithm.
\begin{figure*}
\centering{\includegraphics[width=1.0\textwidth,clip=true]{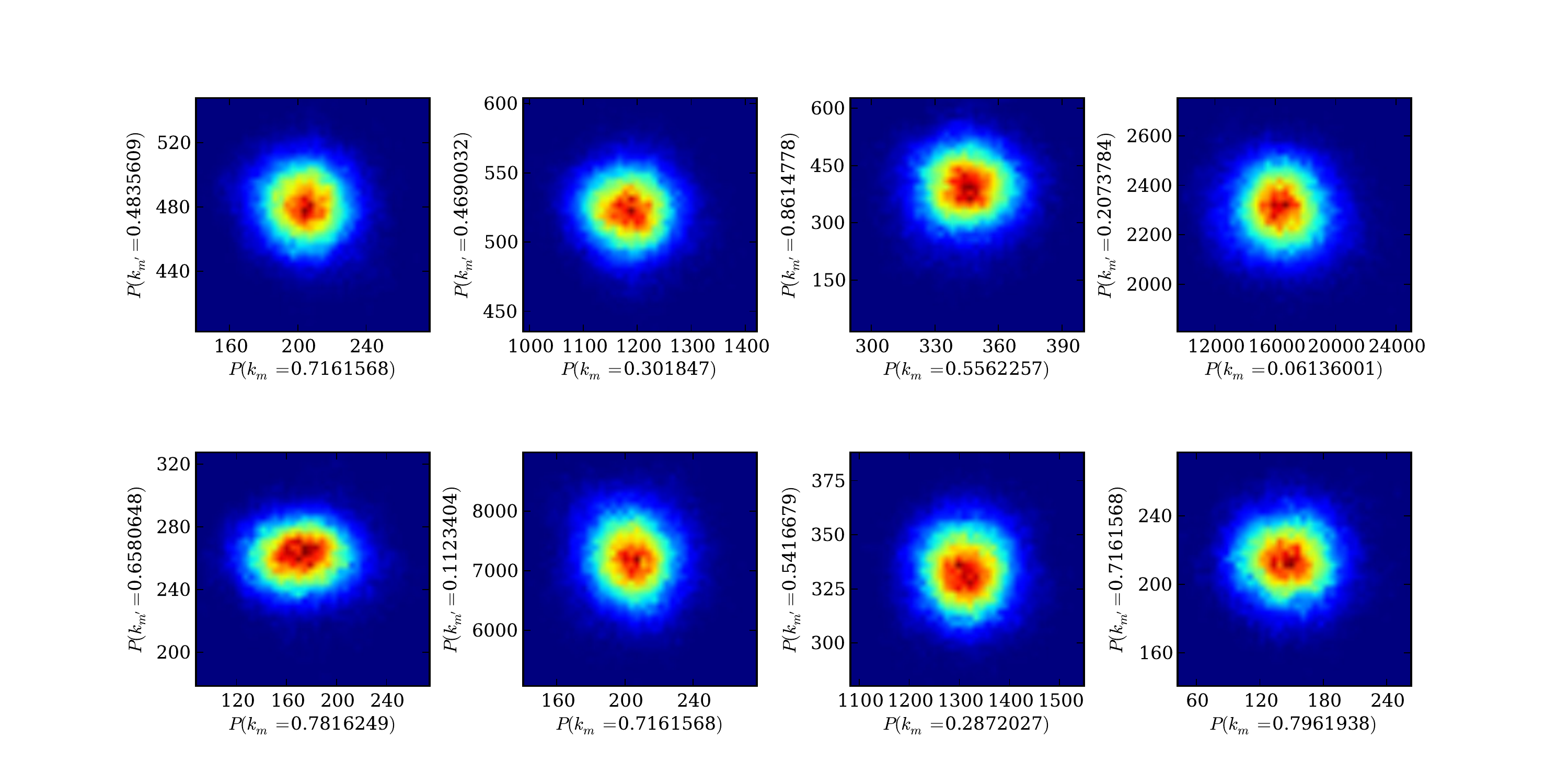}}
\caption{Marginalized two dimensional posterior distribution for randomly selected pairs of power spectrum modes.}
\label{fig:margdens}
\end{figure*}

\section{Generating artificial observations}
\label{mock_observations}
Above we discussed the derivation and numerical implementation of our method.
In this section we will describe the generation of an artificial galaxy survey, used to test our method.
Following closely the description in \citet[][]{JASCHE2010PSPEC}, at first a realization for the density contrast \(\delta_i\) is generated from a normal distribution with zero mean and a covariance matrix corresponding to a cosmological power spectrum.
The density field will be stored on a three dimensional equidistant Cartesian grid with \(N_{side}=128\), corresponding to \(N_{vox}=2097152\) volume elements, and a co moving box length of \(L=750 \mathrm{Mpc}\, h^{-1}\).
The cosmological power spectrum for the underlying matter distribution, with baryonic wiggles, is calculated according to the prescription described in \citet{1998ApJ...496..605E} and \citet{1999ApJ...511....5E}.
Further, a standard \(\Lambda\)CDM cosmology with a set of cosmological parameters (\(\Omega_m=0.24\), \(\Omega_{\Lambda}=0.76\), \(\Omega_{b}=0.04\), \(h=0.73\), \(\sigma_8=0.74\), \(n_s=1\) ) is assumed.

On top of this three dimensional density contrast realization \(\delta_i\), \(M^{\mathrm{tracer}}=15\) artificial galaxy sub samples \(N^l_i\) will be generated corresponding to the data model described in section \ref{datamodel_likelihood}.
For each subsample the noise term \(\epsilon_i^l\) is generated from a multivariate normal distribution with zero mean and a diagonal covariance matrix \(L^l_{ij}=\bar{N}^l\, R^l_i \, \delta^K_{ij}\).

\begin{figure*}
\centering{\includegraphics[width=1.0\textwidth,clip=true]{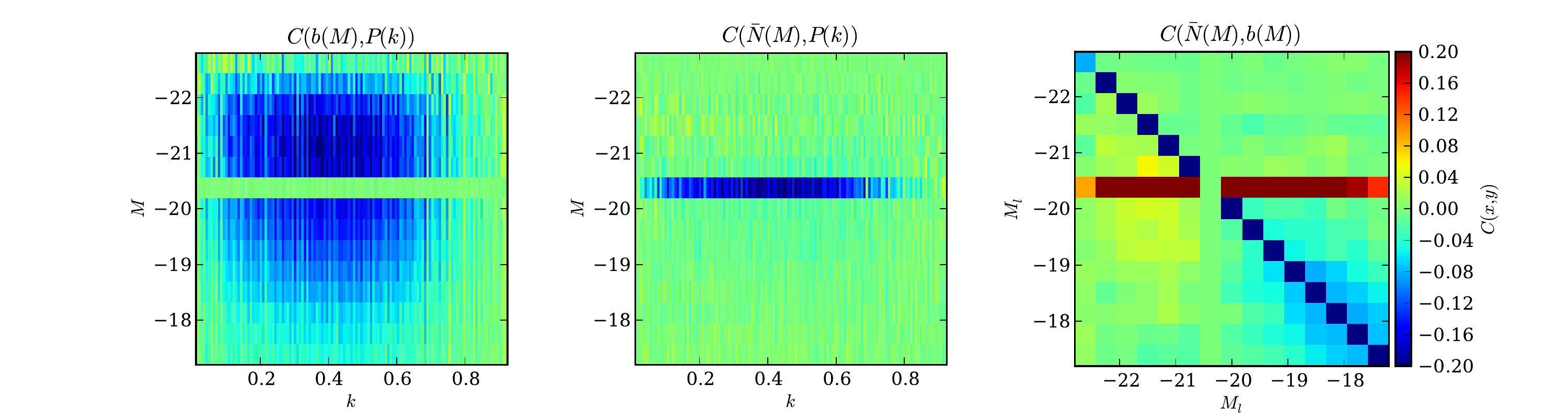}}
\caption{Posterior correlation matrices between the bias parameters \(b_l\) and the power spectrum \(P(k)\) (left panel), between \(\bar{N}_l\) parameters and the power spectrum \(P(k)\) (middle panel), and between the \(b_l\) and \(\bar{N}_l\) parameters (right panel).
In particular, the anti-correlation between inferred bias parameters and the power spectrum at different scales can be clearly seen.  }
\label{fig:mutualcorrmat}
\end{figure*}

In particular, the aim of the present work is to test our method in a realistic scenario with largely inhomogeneous galaxy samples with respect to their clustering behavior.
As already discussed above, such a sample would not represent a spatially homogeneous sample of the three dimensional density field, and hence introduces scale dependent biases in Fourier space \citep[][]{TEGMARK2004}.
 This situation is usually faced when analyzing real galaxy observations \citep[see e.g.][]{TEGMARK2004}.
Subdividing the galaxy sample into \(M^{\mathrm{tracer}}=15\) sub samples permits to account for the respective systematic effects while jointly extracting information on the cosmological power spectrum from a galaxy survey. 

In setting up a realistic test scenario, we emulate the main features of the Sloan Digital Sky survey data release 7 as closely as possible \citep[][]{SDSS7}.
In particular, we use the completeness mask of the Sloan Digital Sky Survey data release 7 depicted in the right panel of figure \ref{fig:selfunc_surveygeometry}. This mask was evaluated with the MANGLE code provided by \citet{SWANSON2008MNRAS} and has been stored on a HEALPIX map with \(n_{side}=4096\) \citep[][]{HEALPIX}. Further, we assume that the radial selection function follows from a standard Schechter luminosity function with standard r-band parameters ( \(\alpha = -1.05\), \(M_* -5 \mathrm{log}_{10}(h)=-20.44\) ), and we only include
galaxies within an apparent Petrosian r-band magnitude range  \(14.5\,<\,r<\,17.77\) and within the absolute magnitude ranges \(M_{min}=-17\) to \(M_{max}=-23\).
The corresponding radial selection function \(f(z)\) is then obtained by integrating the Schechter luminosity function over the range in absolute magnitude:
\begin{equation}
f(z)=\frac{\int^{M_{max}(z)}_{M_{min}(z)} \Phi(M) \, \mathrm{d}M}{ \int^{M_{max}}_{M_{min}} \Phi(M) \, \mathrm{d}M}\, ,
\end{equation}
where \(\Phi(M)\) is given in appendix \ref{schechter_function}.
The selection functions for the galaxy sub samples are depicted in the left panel of figure \ref{fig:selfunc_surveygeometry}.
The product of the survey geometry and the selection function at each point in the three dimensional volume yields the survey response operator:
\begin{equation}
R^l_i = C^l_i\, F^l_i= C(\alpha_i,\delta_i) f^l(z_i) \, ,
\end{equation}
where \(\alpha_i\) and \(\delta_i\) are the right ascension and declination coordinates corresponding to the \(i\)th volume element, and \(z_i\) is the corresponding redshift. \newtext{The functions \(C(\alpha_i,\delta_i)\) and \(f^l(z_i)\) are the continuous survey completeness and selection function, respectively,  sampled at the \(i\){\it{th}} grid position.  }
Further, we will model a luminosity dependent bias,\newtext{ in terms of absolute magnitudes M}, as:
\begin{equation}
b^l=b(M^l)=b_{*}\left( a + b\,10^{0.4\left(M_{*}-M^l \right)} + c\,\left(M^l-M_{*}\right)\right) \, ,
\label{eq:rel_bias}
\end{equation}
with the fitting parameters \(a=0.895\), \(b=0.105\) and \(c=-0.040\) \citep[see ][ for details]{TEGMARK2004}. Since equation (\ref{eq:rel_bias}) formulates only the relative bias with respect to a reference point
in absolute magnitude \(M_{*}\), we will arbitrarily set \(M_{*}=-20.83\) and \(b_{*}=1.2\) for our test scenario.
Finally we have to define a artificial set of mean galaxy numbers \(\bar{N}^l\). In the case studied here, the different \(\bar{N}^l\) are directly related to the Schechter luminosity function by integrating \newtext{over the width of the \(l\){\it{th}} absolute magnitude bin}:
\begin{equation}
\bar{N}^l = \int_{M_{l}}^{M_{l+1}} \Phi(M)\, \mathrm{d}M 
\end{equation}
The individual galaxy observations \(N_i^l\) are then obtained by evaluating equation \ref{eq:data_model} with the quantities as specified above.

\section{Testing the algorithm}
\label{TESTING}
The aim of this section is to conduct a series of tests to judge the performance of our method in a realistic scenario by applying the algorithm to
the simulated galaxy survey described in the previous section. We particularly focus on studying the burn-in and convergence behavior of our method,
 which determines its  overall numerical feasibility and efficiency  in real-world applications.

\subsection{Testing burn-in and statistical efficiency}
\label{convergence}
Metropolis Hastings samplers are designed to have the target distribution  as their stationary distribution \citep[see e.g.][]{metroplis,hastings,NEAL1993}.
The sampling process will provide samples from the specified joint large scale structure posterior distribution after a sufficiently long burn-in phase.
The theory of Metropolis Hastings sampling by itself does not provide any criterion to determine the length of the burn-in phase, so we study it in numerical experiments.

Typically, burn-in manifests itself as a systematic drift of sampled parameter values towards their true underlying values from which the artificial data was generated.
This behavior can be monitored by following the evolution of parameters in subsequent samples \citep[see e.g.][]{2004ApJS..155..227E,JASCHE2010PSPEC}. 
To test this initial burn-in behavior we will arbitrarily set the initial values for the parameters \(\bar{N}^l\) and \(b^l\) to unity \(\bar{N}^l=b^l=1\), while the initial 
power spectrum amplitudes for \(P(k)\) are assumed to be two orders of magnitude larger than their true underlying values. One can then observe the successive drift towards the true values of the respective parameters.
The 1000 first successive samples of the Markov chain are presented in figure \ref{fig:burnin}, which clearly demonstrates the drift towards the respective true solutions.

Figure \ref{fig:burnin} already illustrates the interplay between  bias and power spectrum  samples. 
While bias parameters approach their true values from the top, the power spectrum approaches its fiducial value from below. 
This effect originates from the anti-correlation between biases and power-spectra for given data. In particular high bias values will result in lower power spectrum amplitudes and vice versa.
\begin{figure*}
\centering{\includegraphics[width=1.0\textwidth,clip=true]{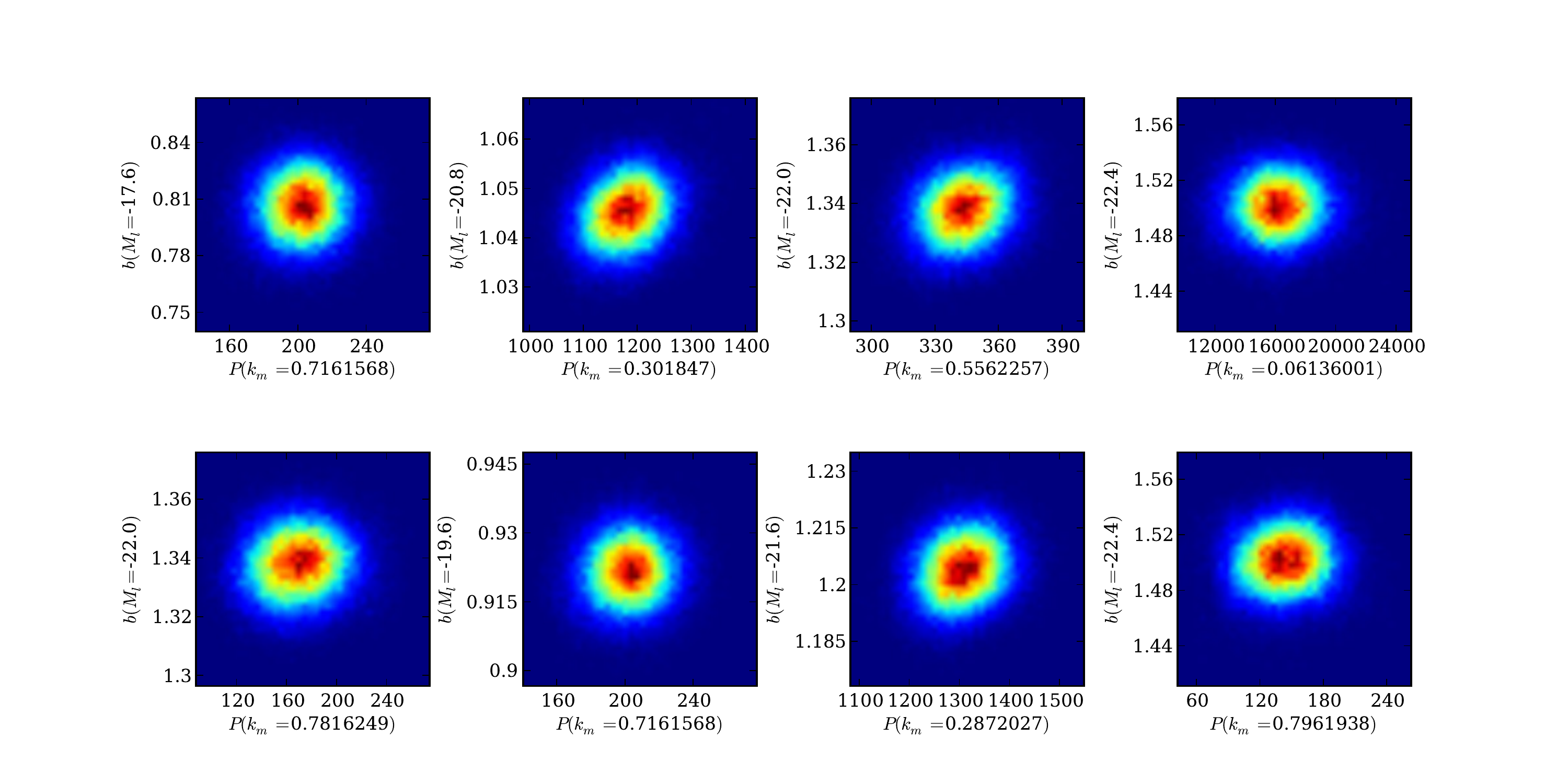}}
\caption{Marginalized two dimensional posterior distributions for randomly selected bias parameters \(b_l\) and power spectrum amplitudes \(P(k_l)\).}
\label{fig:margdens_b}
\end{figure*}

More quantitatively the initial burn-in behavior is studied by comparing the \(s\)th sample of a parameter \(\omega_s\) in the chain to its true underlying value \(\omega^{0}\) via:
\begin{equation}
\label{eq:BURN_IN}
\xi\left(\omega_s\right) = 1-\frac{\omega_s}{\omega^{0}}\, ,
\end{equation}
where for our case, the parameter sample \(\omega_s\) has to be replaced by one of the parameters sampled by the Markov chain.
The results for the \(\xi\left(\bar{N}^l_s\right)\), \(\xi\left(b^l_s\right)\) and \(\xi\left(P_s(k)\right)\) are shown in the lower panels of figure \ref{fig:burnin}.
It can be seen that the algorithm drifts towards the preferred region in parameter space and starts exploring the parameters corresponding to their joint uncertainties.
Also note, that figure \ref{fig:burnin} already indicates the degree of scatter which can be expected in the final analysis of the Markov run.

In this test we do not observe any particular hysteresis for the poorly constrained large scale modes, meaning they do not remain at their initially set values but efficiently explore the parameter space. 
This reflects the ability of our algorithm to account for artificial mode coupling as introduced by the survey geometry, selection effects and biases.

These tests show that the algorithm requires an initial phase of on the order of  \(1000\) samples before providing typical samples from the joint posterior.
This initial burn-in period can be shortened by initializing the algorithm with a set of parameters closer to the true solutions, contrary to the test presented here where the initialization was intentionally chosen to test the burn-in behavior. 

Statistical efficiency is a major design goal of Markov Chain Monte Carlo algorithms. Generally, 
successive samples in the chain are not independent but correlated with their predecessors. Consequently the degree of correlation between successive samples
determines the amount of independent samples which can be drawn from a chain with given length.
To study this correlation effect we follow a similar approach as described in \citep[][]{2004ApJS..155..227E} or \citep[][]{JASCHE2010PSPEC}.

Assuming all parameters in the Markov chain to be independent of each other, the correlation between subsequent samples \(\omega_s\) for the parameter \(\omega\) can be quantified in terms of the  autocorrelation function
\begin{equation}
\label{eq:CORR_COEFF}
C(\omega)_n =\left \langle  \frac{\omega^s-\left \langle \omega\right \rangle}{\sqrt{\mathrm{Var} \left(\omega\right)}} \frac{\omega^{s+n}-\left \langle \omega\right \rangle}{\sqrt{\mathrm{Var} \left(\omega\right)}} \right \rangle \, ,
\end{equation}
where \(n\) is the distance in the chain measured in iterations \citep[also see e.g.][ for a similar discussion]{2004ApJS..155..227E,JASCHE2010PSPEC}.
The results for this analysis are presented in figure \ref{fig:corrlength} which shows the auto-correlation functions for the parameters \(\bar{N}^l\), \(b^l\) and \(P(k)\).

One can  define a correlation length of the Markov sampler as the distance in the chain \(n_c\) beyond which the correlation coefficient \(C(\omega)_n\) has dropped below a threshold of \(0.1\). 
As can be seen in figure \ref{fig:corrlength} the correlation length is typically less than 200 samples, except for a few power-spectrum modes in the Nyquist regime of the box, which are correlated across a few hundred samples.
This demonstrates the overall high statistical efficiency of the sampler.
Note, that these Nyquist modes are under-sampled due to the finite resolution of the Fast Fourier Transform. Even though our method treats these eight Nyquist frequencies correctly, they will not be considered for a typical cosmological analyses where usually only modes with frequencies less than seventy percent of the Nyquist frequency are considered \citep[see e.g.][]{Jing2005}. 

These tests demonstrate the numerical feasibility of our method to explore the full joint posterior distribution despite the high dimensionality
of the problem presented here.

\subsection{Joint uncertainties and correction of systematics}
\label{joint_uncertainties}
The previous section demonstrates the numerical and statistical efficiency of our algorithm to explore the joint large scale structure posterior in high dimensional spaces.
As will be shown in this section, the algorithm also accurately accounts for systematics and joint stochastic uncertainties in the inference of the different parameter.
Of particular importance for the measurement of cosmological parameters from power-spectra are the effects of survey geometry and selection effects. Survey geometry and selection effects introduce artificial mode coupling to the Fourier modes of the three-dimensional density field, and can greatly reduce the visibility
of  features in the power spectrum, such as the baryonic acoustic oscillations \citep[][]{COLE2005}. Correcting these effects is hence of utmost importance for any
method to estimate the cosmological power spectrum from observations.

The information on how well our method corrects for survey geometry effects is naturally contained in the correlation structure of the individual power spectrum samples.
To examine this effect, we calculate the correlation matrix of the power spectrum samples 
\begin{equation}
\label{eq:CORR_COEFF}
\mat{C}_{m\,m'} =\left \langle  \frac{P_m-\left \langle P_m\right \rangle}{\sqrt{Var P_m}} \frac{P_{m'}-\left \langle P_{m'}\right \rangle}{\sqrt{Var P_{m'}}} \right \rangle \, ,
\end{equation}
where \(P_m=P(k_m)\) is the power spectrum amplitude at the mode \(k_m\). 
The result of this exercise is presented in figure \ref{fig:corrmat}, where the ensemble average has been taken over \(7\times10^4\) power spectrum samples.
As can be clearly seen, the resultant correlation matrix exhibits a well defined diagonal structure, as expected from theory. 
Also note the slight red band of anti-correlation around the diagonal, particularly at the highest frequencies in figure \ref{fig:corrmat}.
This anti-correlation indicates that the power spectrum frequency resolution is higher than supported by the data. Note that these effects are accurately accounted for by
our method, since it provides the full probability distribution for each mode. If one wishes, it is possible to reduce frequency resolution in a post-processing step
until anti-correlation vanishes \citep{JASCHE2010PSPEC}.

It should be noted, that generally the posterior distributions for the individual \(P_m\) are non-Gaussian.
Consequently, two-point statistics of the power spectrum, as presented in figure \ref{fig:corrmat}, do not provide conclusive information on the full statistical behavior of the \(P_m\).
Since our method provides a numerical representation of the large scale structure posterior higher order statistics for the power spectrum can naturally be taken into account.
To illustrate this fact, we plot two dimensional marginalized posterior distributions for a few arbitrarily chosen pairs of \(P_m\)s in figure \ref{fig:margdens}.

\begin{figure*}
\centering{\includegraphics[width=1.\textwidth,clip=true]{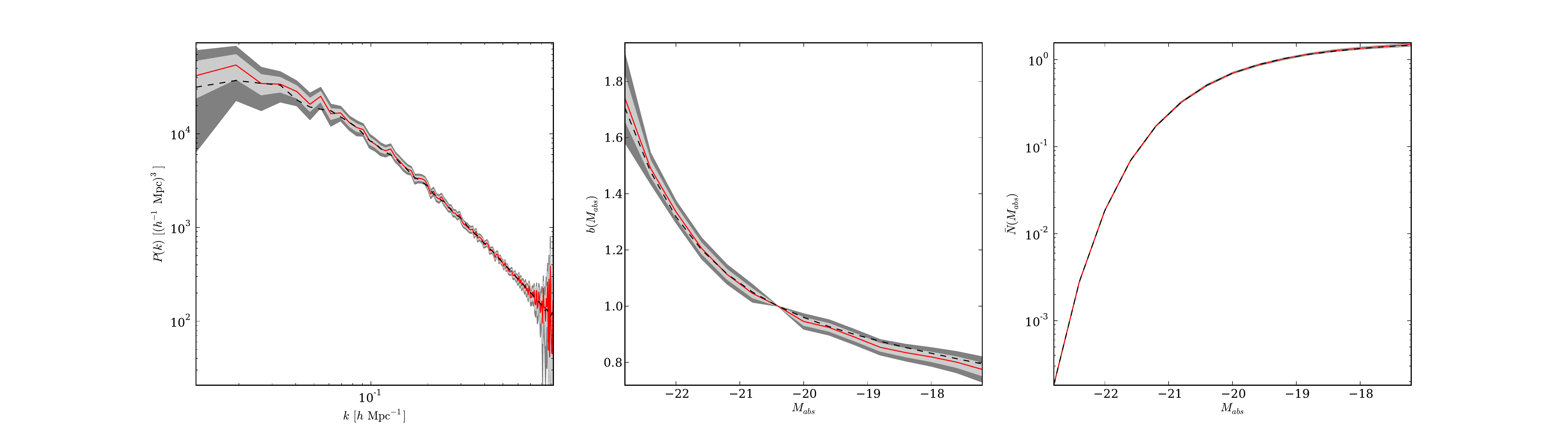}}
\caption{Inferred ensemble means (red lines) for the power spectrum (left panel), bias (middle panel) and \(\bar{N}_l\) (right panel) and corresponding one sigma (light gray shaded) and two sigma (dark gray shaded) credible regions. }
\label{fig:ensemblemeans}
\end{figure*}


In the previous section \ref{convergence} we saw burn-in behavior that suggested  bias parameters \(b_l\)
being anti-correlated with inferred power-spectra \(P(k)\). This behavior is easily understood:  a higher bias parameter requires a lower amplitude of the power spectrum to match observations. The details and strength of these effects generally
depend on the magnitude bin for the bias parameter and the mode of the power spectrum. To quantify these effects we now study the posterior correlations
between the inferred bias parameters \(b_l\), the \(\bar{N}_l\) and the power spectrum \(P(k)\).  Figure
\ref{fig:mutualcorrmat} shows the  correlation matrices 
\begin{equation}
\mat{C}(x,y) =\left \langle  \frac{x-\left \langle x\right \rangle}{\sqrt{Var\, x}} \frac{y-\left \langle y\right \rangle}{\sqrt{Var\, y}} \right \rangle \, ,
\end{equation}
where \(x\) and \(y\) have to be replaced by the parameter under consideration.
The scale dependent anti-correlation between bias parameters and power-spectra is clearly visible in the left panel of figure \ref{fig:mutualcorrmat}.
Note, that we are measuring relative biases and for that reason the bias amplitude at a magnitude \(M_l=-20.4\) is fixed to 1.

The maximum anti-correlation amounts to about 20 per cent, and significant anti-correlation can be observed at all
scales and for all magnitudes. This result emphasizes our claim, that even simple linear bias amplitudes cannot be estimated independently from power-spectra and vice versa.

The middle panel of figure \ref{fig:mutualcorrmat} shows the correlations between the \(\bar{N}_l\) and different modes of the power spectrum.
Generally correlations and anti-correlations amount maximally up to a few percent. However, \(\bar{N}_l\) at the magnitude bin
\(M_l=-20.4\) of the fixed bias amplitude is  strongly anti-correlated with the modes of the power spectrum. This indicates, that \(\bar{N}_l\)
carries additional information on the bias amplitude at the magnitude bin \(M_l=-20.4\). 

Finally, the right panel of figure \ref{fig:mutualcorrmat} shows the correlations between
inferred \(\bar{N}_l\) and  \(b_l\) parameters. The prominent diagonal character of this correlation matrix demonstrates that 
the anti-correlation between inferred biases and \(\bar{N}_l\) is strongest for values belonging to the same magnitude bin. Nevertheless, there exist percent level 
correlations and anti-correlations between parameters in different bins.  A prominent feature of this plot is also the strong band of positive correlations
for the parameter \(\bar{N}_l\) at magnitude \(M_l=-20.4\), which demonstrates strong correlations between \(\bar{N}_l\) in this bin and the bias parameters \(b_l\) of other magnitude bins.


Since our method
provides a numerical representation of the highly non-Gaussian posterior distribution, 
uncertainty quantification is not limited to Gaussian approximation but can take into account all involved non-Gaussianities
and non-linearities. As a demonstration, in figure \ref{fig:margdens_b}, we plot marginalized two dimensional distributions of individual, randomly selected, bias parameters \(b_l\)
and modes of the power spectrum \(P(k)\). It can be seen, that the marginalized distributions reflect the previously observed anti-correlations between
bias parameter and power-spectra. Also note, that the marginalized distributions, presented in figure \ref{fig:margdens_b}, exhibit non-Gaussian behavior. 
\newtext{The results of this section support the requirement for a full joint approach to cosmological power-spectrum estimation.}


\begin{figure*}
\centering{\includegraphics[width=1.0\textwidth,clip=true]{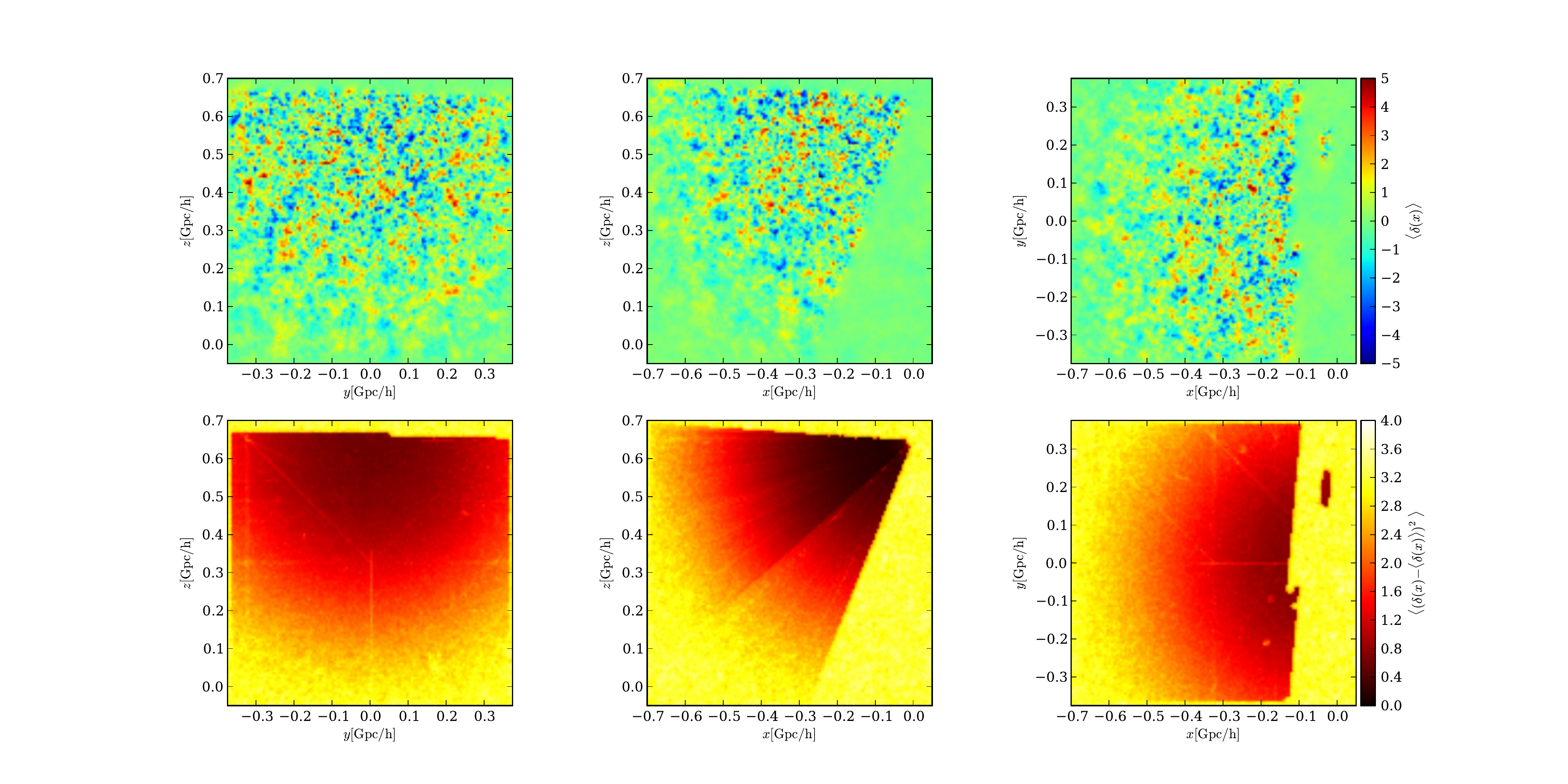}}
\caption{Three slices from different directions through the three dimensional posterior mean density (upper panels) and standard deviation (lower panels) fields. It can be seen, that the method accounts for
non-trivial survey geometry. 
}
\label{fig:ensemblemeans_dens}
\end{figure*}

\section{Results}
\label{results}
In the previous section we discussed the numerical feasibility of our method and demonstrated the necessity of a full joint approach.
Here we will describe the final results inferred from the Markov chain. 
As described in section \ref{mock_observations}, we seek to test our method in a realistic scenario by emulating characteristic features of the
SDSS DR7 main sample. A galaxy sample, such as the SDSS main sample, represents an interesting and challenging test scenario for our method, since it
consists of a large class of galaxy types exhibiting different clustering behavior at different redshifts and hence requires to exploit
the entire methodology as described in this work. On the other hand, the SDSS main sample is a shallow galaxy sample with a median redshift of about \(z_{med}=0.1\),
which is not ideal to infer tiny features at large scales in the power spectrum, such as the baryon acoustic oscillations. 

Additional information, to probe the large scales,
can be obtained from highly clustered luminous red galaxies (LRG) which cover larger volumes to higher redshifts. For the sake
of demonstrating the capabilities of our code to  infer the power spectrum
 on the \textit{entire} range of scales covered by the survey, we ignored LRGs in this work.
Note, that LRGs can be easily added as an additional
data bin in the joint analysis, where they would add their information content to the inference process. 
It should be remarked, that  the analysis volume of \( 750^3 \, \rm{Mpc^3/h^3}\), used for the inference, is only about half filled by data..
This obviously increases the challenge compared to a case where the empty volume would be filled with additional LRGs.
In this respect, the presented test is harder compared to a real cosmological analysis in which we would exploit all available information.

To summarize the inferred quantities, in figure \ref{fig:ensemblemeans}, we show ensemble means for the power spectrum, the bias and \(\bar{N}_l\), estimated from \(72\times 10^4\)
Markov realizations. It is immediately obvious, that the mean density parameters \(\bar{N}_l\) can be recovered very well.  The middle panel of figure \ref{fig:ensemblemeans}
depicts the ensemble mean for the bias parameters \(b_l\) and corresponding uncertainties. The plot demonstrates that 
the shape of the magnitude dependent bias function can be accurately recovered from data in this joint analysis. Since we are measuring relative biases the bias parameter at a magnitude of \(M_l=-20.4\) is held fixed at unity.
  
The left panel of figure \ref{fig:ensemblemeans} depicts the ensemble mean power spectrum and corresponding uncertainties.
The inferred ensemble mean power spectrum nicely follows the true underlying fiducial power spectrum across
the entire range of modes covered by our analysis.  Even on the largest scales,  poorly constrained by the
artificial galaxy survey dominated by sample variance, the analysis returns sensible constraints and hints at the turn-over in the power spectrum.

Visually it is not possible to observe recovered baryon acoustic oscillations. This is due to several effects.
First of all, the analyzed volume is too shallow to accurately probe the Baryon acoustic oscillations.  Secondly,
joint inference of the power spectrum, \(b_l\) and \(\bar{N}_l\) correctly accounts for  coupled uncertainties  which will 
lead to a blurring of tiny features in the power spectrum compared to an analysis that takes best-fit biases as inputs to power spectrum analysis. We conclude that our artificial data set, emulating the SDSS main sample data, is not sufficiently informative
to provide conclusive information on  the baryon acoustic oscillations. Although preliminary tests with real observations already indicate 
qualitatively similar behavior, a detailed analysis of the latest SDSS data will be subject to a forthcoming publication.  

Finally, we would like to point out, that our method also provides measurements of the three dimensional density field
and corresponding uncertainties. Individual samples of these density fields represent constrained realizations of the underlying density field cleaned from all statistical and systematic uncertainties,
and as such will generally be valuable to study galaxy and three dimensional large scale structure formation. Note that in contrast to conventional constrained realizations  our method obtains these reconstructions without assuming biases, number densities, or correlations \textit{a priori}.

The slices through the posterior mean density shown in figure \ref{fig:ensemblemeans_dens} can be thought of as a non-linear Wiener filter. Our method also delivers an error model for this reconstruction, here summarized in terms of standard deviation per pixel. These plots illustrate that our method is able to account for arbitrary and complex
survey geometries in three dimensional space.

\section{summary and conclusion}
\label{Conclusion}
This work describes the derivation and numerical implementation of a full Bayesian approach to cosmological large scale structure power spectrum analysis.

It addresses the problem of joint power spectrum and galaxy bias inference from redshift surveys, and thus aims at a natural and fully self consistent treatment of joint and correlated uncertainties.
Traditionally, these correlations remain generally unaccounted for in standard sequential pipeline approaches to power spectrum inference, which assume fixed bias models \citep[see e.g.][]{PERCIVAL2004,JASCHE2010PSPEC}. 
The method, as described in this work, naturally accounts for correlated uncertainties between all quantities to be inferred, by exploring the joint parameter space, of the three dimensional density field, the power spectrum, luminosity dependent
biases and corresponding normalizations. This is achieved through the use of multiple block Metropolis Hastings and Hybrid Monte Carlo algorithms.

As described in section \ref{LSS_POSTERIOR},  the multiple block sampling approach permits us to dissect the computationally challenging task of joint sampling into sequential
sub tasks of lesser complexity. Thus, the problem can be processed as outlined in figure \ref{fig:flowchart}, yielding a numerical representation of the full joint posterior distribution.

The algorithm presented in this work improves over the method presented in \citet{JASCHE2010PSPEC}, in various respects. Most notably, the previous three dimensional density sampling procedure, relying
on the solution of large systems of linear equations via Krylov space methods, has been replaced by a Hybrid Monte Carlo scheme. This has proven to be  more efficient, especially in high dimensions. 

We further increased the statistical efficiency
and hence the number of independent samples which can be drawn from the Markov chain. To do this, we followed the deterministic transition approach proposed by \citet{JEWELL2009},
but incorporated an HMC transition, rather than a pure Metropolis-Hastings step, to improve the overall computational efficiency of the algorithm.  

Further, we derived conditional posterior distributions for the inference of absolute magnitude dependent bias and corresponding normalizations. As described in section \ref{LSS_POSTERIOR},
we followed a maximally agnostic approach by assuming all absolute magnitude bins to be independent and additionally assuming flat priors for the corresponding bias and normalization amplitudes.
The resultant conditional posterior distributions have analytic expressions. In particular, realizations of magnitude dependent biases can be generated from univariate normal distributions and corresponding normalization factors are drawn from generalized inverse Gaussian distributions.

We also described tests of our method in a realistic scenario. Section \ref{mock_observations} describes the generation of an artificial
galaxy survey emulating characteristic features of the SDSS DR7, such as survey geometry, selection effects, and luminosity dependent galaxy biases.
The test was particularly designed to highlight the problematics of survey geometry, selection effects and luminosity dependent biases and their impact on
cosmological power spectrum inference.
Specifically, the main goals of the test, described in section \ref{TESTING}, is to study the expected numerical behavior and efficiency of our method when applied to real data.
Through a simple numerical experiment, by setting the initial values of the bias and normalization factors to unity and scaling the amplitude of the initial power spectrum up by two orders
of magnitude, we were able to determine the initial burn-in phase of the Markov sampler to be on the order of 1000 samples, which demonstrates numerical feasibility of the method.
Also note, that initial burn-in time can be significantly shortened by a more realistic choice of the initial parameter values.

We overcame a crucial difficulty in implementing a joint inference approach: statistical dependency between the bias parameters and the power spectrum leads to unfeasibly long correlation lengths for our block sampling scheme. Adding a deterministic transition step reduces the  correlation length of typical parameters by a factor of more than 1000.
As a remark, in tests without such a deterministic transitions we observed still eighty percent correlations at a lag of 10000 samples for most of the power-spectrum modes yielding an unfeasibly long  convergence time of the sampler.
This dramatic improvement allows practical use of our algorithm even in regimes where signal-to-noise is low.

Our method accurately accounts for artificial mode coupling due to the survey geometry and selection effects.
The correlation structure of the Markov samples in Fourier space reveals that the power spectrum correlation matrix is of highly diagonal shape.
Even when coupled with inference of the bias parameters, maximal residual correlations between power spectrum bins are typically less than 10 per cent, emphasizing the ability of our method to account for survey geometry effects.

Our test revealed anti-correlations at the level of 20 per cent between the bias parameters and the power spectrum across large ranges in Fourier-space. 
 It remains to be studied to what extent this coupling  has impacted past cosmological analyses of galaxy survey data that ignored it. 

While a linear luminosity dependent galaxy bias is one of the simplest bias models for cosmological inference from galaxy surveys, it does not seem reasonable to assume that  more complicated bias models would not suffer from this coupling effect. 
Still, our method is easily extensible to include more complicated bias models such as  second and higher order bias.

In section \ref{results}, we summarize means and credible regions for the parameters inferred by our method. In particular, the power spectrum
can be accurately recovered from the test data. 

Since we treat each luminosity class as a separate galaxy sample with its own bias parameters,  we actually inferred the power spectrum from 15 different galaxy sub-samples all exhibiting different
systematic and statistical uncertainties. In particular, each sub-sample comes with its own mask,  average number density, bias, and selection function.  Our method is therefore able to perform joint power spectrum analysis from combinations of different galaxy surveys---and, after further development, possibly even different probes of large scale structure.

Note, that the normalization parameters in the different magnitude bins are directly related to the luminosity function, which in a scenario
as described in this work, is a natural byproduct of the cosmological inference. Also note, that our method further provides full three-dimensional maps of the inferred
matter distributions and corresponding uncertainties. These constrained realizations of the density distribution may provide useful information to test
cosmological structure formation.

The method, as described here, seamlessly integrates into the larger  Bayesian large scale structure inference framework.
In particular, it can be used in conjunction with the photometric redshift sampling method described in \citet{JASCHEPHOTOZ}, to account for additional 
redshift uncertainties. In that combination our method would accurately treat all joint and correlated uncertainties when analyzing galaxy surveys with
highly uncertain radial positions of galaxies. \newtext{Note, although not discussed in this work, the present bias sampling framework seamlessly integrates with effective measures
of redshift space distortions corrections, for instance as described in \citet{TEGMARK2004}, \citet{2006MNRAS.373...45E} or \citet{JASCHE2010HADESDATA}, but also with information theoretically more rigorous approaches, which will be subject to future publications. }

\newtext{In addition,} a exciting next step will be to merge our method with the initial conditions inference framework presented in \citep{JASCHE2012}. This would permit inferring the power spectrum of the initial conditions of the Universe from catalogs of biased tracers.

In summary, our method jointly infers all parameters  in the cosmological power spectrum inference problem and
 accurately accounts for correlated uncertainties and interdependencies. As a highly flexible and numerically efficient approach it  seamlessly integrates into a larger body of 
large scale structure inference methods and allows straightforward generalization to more sophisticated biasing schemes.
The application of this approach to actual data will be the subject of a forthcoming publication. Preliminary tests  indicate a qualitatively similar behavior to the results presented here. 

\section*{Acknowledgments}
We thank Andr\'es Balaguera Antolinez and Franz Elsner for useful discussions and support in the course of this project. 
JJ is partially supported by a Feodor Lynen Fellowship by the Alexander von Humboldt foundation. BDW acknowledges  support from NSF grants AST 07-08849 and AST 09-08693 ARRA, and a Chaire d'Excellence from the Agence Nationale de Recherche and computational resources provided through  XSEDE grant AST100029. This material is based upon work supported in part by the National Science Foundation under Grant No. PHYS-1066293 and the hospitality of the Aspen Center for Physics.

\bibliography{paper}
\bibliographystyle{apj}


\appendix
\section{Inverse Gaussian distribution for \(\bar{N}^l\)}
\label{inv_gauss}
As described in section the posterior distribution for the mean galaxy numbers \(\bar{N}^l\) is proportional to the data likelihood: 
\begin{eqnarray}
\mathcal{P}\left(\bar{N}^l|\{N^l_i\},\{\delta_i\},b^l\right)&\propto&\mathcal{P}^l\left(\{N^l_i\}|\{\delta_i\},\bar{N}^l,b^l\right) \nonumber \\
&\propto& \frac{\mathrm{e}^{-\frac{1}{2}\sum_{i}^{N_{vox}} \frac{\left(N^l_i-\bar{N}^l\,R^l_i\,\left(1+b^l\,\delta_i \right)\right)^2}{\bar{N}^l\,R^l_i}}}{\left(\bar{N}^l\right)^{\frac{N_{vox}}{2} }} \nonumber \\
\label{eq:nmean_post}
\end{eqnarray}
Expanding the square in the exponent and discarding numerical constants then yields:
\begin{eqnarray}
\mathcal{P}\left(\bar{N}^l|\{N^l_i\},\{\delta_i\},b^l\right)&\propto& \frac{\mathrm{e}^{-\frac{1}{2}\sum_{i}^{N_{vox}} \frac{\left(N^l_i\right)^2-2\,N^l_i\,\bar{N}^l\,R^l_i\,\left(1+b^l\,\delta_i \right)+\left(\bar{N}^l\,R^l_i\,\left(1+b^l\,\delta_i \right)\right)^2}{\bar{N}^l\,R^l_i}}}{\left(\bar{N}^l\right)^{\frac{N_{vox}}{2}  }} \nonumber \\
&\propto& \frac{\mathrm{e}^{-\frac{1}{2} \left( \frac{1}{\bar{N}^l}\sum_{i}^{N_{vox}} \frac{\left(N^l_i\right)^2}{R^l_i}  + \bar{N}^l \sum_{i}^{N_{vox}} R^l_i\,\left(1+b^l\,\delta_i \right)^2  \right)}}{\left(\bar{N}^l\right)^{\frac{N_{vox}}{2} } } \nonumber \\
\label{eq:nmean_post}
\end{eqnarray}
The notation can be simplified by introducing the following quantities:
\begin{equation}
A^l=\sum_{i}^{N_{vox}} R^l_i\,\left(1+b^l\,\delta_i \right)^2\, ,
\end{equation}
and
\begin{equation}
B^l=\sum_{i}^{N_{vox}} \frac{\left(N^l_i\right)^2}{R^l_i} \, .
\end{equation}
With these definitions we obtain a one dimensional distribution:
\begin{eqnarray}
\mathcal{P}\left(\bar{N}^l|\{N^l_i\},\{\delta_i\},b^l\right)&\propto& \frac{\mathrm{e}^{-\frac{1}{2} \left( \bar{N}^l A^l+ \frac{1}{\bar{N}^l} B^l  \right)}}{\left(\bar{N}^l\right)^{\frac{N_{vox}}{2} } } \, .
\label{eq:nmean_post_xx}
\end{eqnarray}
It can be seen that equation \ref{eq:nmean_post_xx} describes a generalized inverse Gaussian distribution (GIG). Introducing proper normalization constants then yields:
\begin{eqnarray}
\mathcal{P}\left(\bar{N}^l|\{N^l_i\},\{\delta_i\},b^l\right)= \left(\frac{A^l}{B^l}\right)^{\frac{p^l}{2}} \frac{\left(\bar{N}^l\right)^{p^l-1}}{2K_{p^l}\left(\sqrt{A^l\,B^l}\right)}\,\mathrm{e}^{-\frac{1}{2} \left( \bar{N}^l A^l+ \frac{1}{\bar{N}^l} B^l  \right)} \, \,
\label{eq:nmean_post_xy}
\end{eqnarray}
where \(K_{p^l}(x)\) is a modified Bessel function of the second kind and \(p^l=1-N_{vox}/2\). Generating the mean number of galaxies \(\bar{N}^l\) from this generalized inverse Gaussian
distribution is straightforward.

\section{Normal distribution for the bias}
\label{bias_gauss}
As described in section \ref{sample_the_bias} the conditional distribution for the galaxy bias is given up to a normalization constant as: 
\begin{eqnarray}
\mathcal{P}\left(b^l|\{\delta_i\},\bar{N}^l,\{N^l_i\}\right) &\propto& \mathrm{e}^{-\frac{1}{2}\sum_{i}^{N_{vox}} \frac{\left(N^l_i-\bar{N}^l\,R^l_i\,\left(1+b^l\,\delta_i \right)\right)^2}{\bar{N}^l\,R^l_i}} \nonumber \\
\end{eqnarray}  
By expanding the square in the exponent and discarding numerical constants we obtain:
\begin{eqnarray}
\mathcal{P}\left(b^l|\{\delta_i\},\bar{N}^l,\{N^l_i\}\right) &\propto& \mathrm{e}^{-\frac{1}{2} \left( \left(b^l\right)^2 \sum_{i}^{N_{vox}} \bar{N}^l\,R^l_i\,\delta_i^2 - 2\,b^l \sum_{i}^{N_{vox}} \left(N^l_i-\bar{N}^l\right)\,\delta_i   \right)} \nonumber \\
\end{eqnarray}  
We can simplify notation considerably by introducing the following quantities: 
\begin{equation}
C^l=\sum_{i}^{N_{vox}} \bar{N}^l\,R^l_i\,\delta_i^2 \, ,
\end{equation}
and
\begin{equation}
D^l=\sum_{i}^{N_{vox}} \left(N^l_i-\bar{N}^l\right)\,\delta_i  \, ,
\end{equation}
With this notation we obtain:
\begin{eqnarray}
\mathcal{P}\left(b^l|\{\delta_i\},\bar{N}^l,\{N^l_i\}\right) &\propto& \mathrm{e}^{-\frac{1}{2} \left( \left(b^l\right)^2 C^l - 2\,b^l D^l   \right)} \nonumber \\
&\propto& \mathrm{e}^{-\frac{1}{2} C^l \left( b^l - \frac{D^l}{C^l}\right)^2} \, ,\nonumber \\
\end{eqnarray}  
where in the second line the square in the exponent has been completed. By introducing a variance as:
\begin{equation}
\left(\sigma_{b^l}\right)^2 = \frac{1}{C^l} \, , 
\end{equation}
and a mean as:
\begin{equation}
\mu_{b^l} =\frac{D^l}{C^l} \, , 
\end{equation}
we obtain a univariate normal distribution for the bias parameter \(b_l\):
\begin{eqnarray}
\mathcal{P}\left(b^l|\{\delta_i\},\bar{N}^l,\{N^l_i\}\right) = \frac{\mathrm{e}^{-\frac{1}{2} \frac{ \left( b^l - \mu_{b^l} \right)^2 }{\left(\sigma_{b^l}\right)^2}}}{\sqrt{2\,\pi \left(\sigma_{b^l}\right)^2} } \, .\nonumber \\
\end{eqnarray} 
This result permits us to employ simple sampling techniques of generating random normal variates to rapidly explore
the parameter space of galaxy biases.

\section{Increasing Statistical efficiency}
\label{mixingefficiency}
Splitting up the galaxy sample into a number of sub samples to treat their individual systematics also yields higher uncertainty, since the 
shot noise for each sub sample is increased due to a decreased number of galaxies compared to the full sample. Hence, in order to improve
the mixing behaviour of the Markov chain, it is desirable that the algorithm performs large moves in the power spectrum in regimes with low signal to noise.
This can be achieved by introducing a deterministic proposal distribution to the Metropolis Hastings algorithm \citep[see ][ for details]{JEWELL2009}.
In particular, \citet{JEWELL2009} proposed the following deterministic step for the density field in Fourier space:
\begin{equation}
\hat{\delta}^{s+1}_k = \sqrt{\frac{P^{s+1}(k)}{P^{s}(k)}} \hat{\delta}^{s}_k \, .
\end{equation}
Given such a deterministic transition, a proposal distribution \(w\left( \{\hat{\delta}^{s+1}_k\} , \{ P^{s+1}(k) \} | \{\hat{\delta}^{s}_k\} , \{ P^{s}(k) \} \right) \) for the joint transition of power-spectra and density fields can be given
as:
\begin{eqnarray}
w\left( \{\hat{\delta}^{s+1}_k\} , \{ P^{s+1}(k) \} | \{\hat{\delta}^{s}_k\} , \{ P^{s}(k) \} \right) &=& w'\left( \{ P^{s+1}(k) \} | \{\hat{\delta}^{s}_k\} , \{ P^{s}(k) \} \right) \nonumber \\
& & \,  w''\left( \{\hat{\delta}^{s+1}_k\}|  \{ P^{s+1}(k) \} , \{\hat{\delta}^{s}_k\} , \{ P^{s}(k) \}\right) \nonumber \\
&=& w'\left( \{ P^{s+1}(k) \} | \{\hat{\delta}^{s}_k\} , \{ P^{s}(k) \} \right) \nonumber \\
& & \, \prod_k \delta^D\left( \hat{\delta}^{s+1}_k -  \sqrt{\frac{P^{s+1}(k)}{P^{s}(k)}} \hat{\delta}^{s}_k \right) \, , \nonumber \\
\end{eqnarray}
where in the last line we made explicit use of the deterministic transition in the density field parameters.
Consequently, the Metropolis Hastings acceptance rule for the joint transition is:
\begin{eqnarray}
\label{eq:MH_STEp_MIXING_1}
A\left(  \{\hat{\delta}^{s+1}_k\} , \{ P^{s+1}(k) \} | \{\hat{\delta}^{s}_k\} , \{ P^{s}(k) \} \right) & = & \mathrm{min}\left[1, \frac{\mathcal{P}\left( \{\hat{\delta}^{s+1}_k\} , \{ P^{s+1}(k) \}| \{d\}\right)}{\mathcal{P}\left( \{\hat{\delta}^{s}_k\} , \{ P^{s}(k) \}| \{d\}\right)} \right . \nonumber \\
& & \frac{w'\left( \{ P^{s}(k) \} | \{\hat{\delta}^{s+1}_k\} , \{ P^{s+1}(k) \} \right)}{w'\left( \{ P^{s+1}(k) \} | \{\hat{\delta}^{s}_k\} , \{ P^{s}(k) \} \right)}  \, \nonumber \\
& &  \prod_k \sqrt{\frac{P^{s+1}(k)}{P^{s}(k)}} \nonumber \\
& & \left . \delta^D\left( \hat{\delta}^{s+1}_k -  \sqrt{\frac{P^{s+1}(k)}{P^{s}(k)}} \hat{\delta}^{s}_k \right) \right ] \, .\nonumber \\
\end{eqnarray}
The Dirac delta distribution in the last line of equation (\ref{eq:MH_STEp_MIXING_1}) permits us to rewrite the Metropolis Hastings acceptance rule as:
\begin{eqnarray}
\label{eq:MH_STEp_MIXING_2}
A\left(  \{\hat{\delta}^{s+1}_k\} , \{ P^{s+1}(k) \} | \{\hat{\delta}^{s}_k\} , \{ P^{s}(k) \} \right) & = & \mathrm{min}\left[1, \frac{\mathcal{P}\left(\{ P^{s+1}(k)\} \right)}{\mathcal{P}\left(\{ P^{s}(k) \}\right)}  \right . \nonumber \\
& & \times \prod_l \frac{\mathrm{e}^{-\frac{1}{2}\sum_{i}^{N_{vox}} \frac{\left(N^l_i-\bar{N}^l\,R^l_i\,\left(1+b^l\,C\sum \sqrt{\frac{P^{s+1}(k)}{P^{s}(k)}} \hat{\delta}^{s}_k e^{2\pi i k \frac{\sqrt{-1}}{N}}  \right)\right)^2}{\bar{N}^l\,R^l_i}}}{\mathrm{e}^{-\frac{1}{2}\sum_{i}^{N_{vox}} \frac{\left(N^l_i-\bar{N}^l\,R^l_i\,\left(1+b^l\,C\sum \sqrt{\frac{P^{s}(k)}{P^{s}(k)}} \hat{\delta}^{i}_k e^{2\pi i k \frac{\sqrt{-1}}{N}}  \right)\right)^2}{\bar{N}^l\,R^l_i}}} \nonumber \\
& & \times \left .\frac{w'\left( \{ P^{s}(k) \} | \{\hat{\delta}^{s}_k\} , \{ P^{s+1}(k) \} \right)}{w'\left( \{ P^{s+1}(k) \} | \{\hat{\delta}^{s}_k\} , \{ P^{s}(k) \} \right)} \right ]  \, . \nonumber \\
\end{eqnarray}
As already pointed out by \citet{JEWELL2009}, the result given in equation (\ref{eq:MH_STEp_MIXING_2}), demonstrates that, given a deterministic transition for the density field,
one has to draw random realizations for the power spectrum from the following distribution:
\begin{eqnarray}
\label{eq:spec_posterior_MIXING_1}
\mathcal{P} \left(\{ P^{s+1}(k) \} | \{u_k\} , \{ d \} \right) & = & \mathcal{P}\left(\{ P^{s+1}(k)\} \right) \, \nonumber \\
&  &  \prod_l \mathrm{e}^{-\frac{1}{2}\sum_{i}^{N_{vox}} \frac{\left(N^l_i-\bar{N}^l\,R^l_i\,\left(1+b^l\,C\sum \sqrt{P^{s+1}(k)} \, \hat{u}_k e^{2\pi i k \frac{\sqrt{-1}}{N}}  \right)\right)^2}{\bar{N}^l\,R^l_i}} \nonumber \\
\end{eqnarray}
where we introduced \(\hat{u}_k= \hat{\delta}^{s}_k / P^{s}(k) \).
Drawing random variates from this distribution is a non-trivial task and will in general require a Metropolis-Hastings sampling framework \citet{JEWELL2009}.
In order to increase sampling efficiency and for consistency with the remainining sampling algorithm we will generally use a Hybrid Monte Carlo algorithm, as described in section \ref{HAMILTONIAN_SAMPLING},
to explore the power spectrum posterior given in equation (\ref{eq:MH_STEp_MIXING_2}).

\subsection{Discrete Fourier transformation}
\label{Discrete_Fourier_transformation}
In this work we employ the following convention for the discrete Fourier transform:
\begin{equation}
\label{eqn:FFT}
y_k=\hat{C}\sum_{j=0}^{N-1}x_je^{-2\pi j k \frac{\sqrt{-1}}{N}} \, ,
\end{equation}
while the inverse Fourier transform is given as:
\begin{equation}
\label{eqn:IFFT}
x_j=C\sum_{k=0}^{N-1}y_k e^{2\pi j k \frac{\sqrt{-1}}{N}} = C\sum_{k=-(\frac{N}{2}-1)}^{\frac{N}{2}}y_k e^{2\pi j k \frac{\sqrt{-1}}{N}}  \, .
\end{equation}
Note, that the normalization coefficients \(C\) and \(\hat{C}\) have to fulfill the requirement:
\begin{equation}
\label{eqn:Normalisation_condition}
C\hat{C}=\frac{1}{N} \, .
\end{equation}

\section{The Schechter Luminosity function}
\label{schechter_function}
According to \citet{SCHECHTER1976} the luminosity function can be expressed as:
\begin{equation}
\Phi(M)\, \mathrm{d}M = 0.4\, \Phi^{*} \mathrm{ln}(10)\, \left(10^{0.4\,(M^{*}-M)}\right)^{\alpha+1}\, \mathrm{e}^{-10^{0.4\,(M^{*}-M)}}\,\mathrm{d}M \, . 
\end{equation}
It should be remarked, that for calculating selection functions the normalization \(\Phi^{*}\) is not required.


\label{lastpage}

\end{document}